\documentclass[12pt]{article}
\usepackage{amsmath,amssymb}
\usepackage{graphicx}
\newcommand{\eqdef}{\stackrel{\text{def}}{=}}
\newcommand{\n}{\nonumber\\}
\newcommand{\bm}{\boldsymbol}
\newcommand{\ignore}[1]{}
\renewcommand{\theequation}{\arabic{section}.\arabic{equation}}

\allowdisplaybreaks[4]

\begin{document}

\baselineskip=20pt

\newfont{\elevenmib}{cmmib10 scaled\magstep1}
\newcommand{\preprint}{
    \begin{flushleft}
      \elevenmib Yukawa\, Institute\, Kyoto\\
    \end{flushleft}\vspace{-1.3cm}
    \begin{flushright}\normalsize \sf
      DPSU-10-1\\
      YITP-10-15\\
    \end{flushright}
    }
\newcommand{\Title}[1]{{\baselineskip=26pt
    \begin{center} \Large \bf #1 \\ \ \\ \end{center}}}
\newcommand{\Author}{\begin{center}
    \large \bf Leonor Garc\'ia-Guti\'errez${}^a$,
     Satoru Odake${}^b$
     and Ryu Sasaki${}^a$\end{center}}
\newcommand{\Address}{\begin{center}
      ${}^a$ Yukawa Institute for Theoretical Physics,\\
      Kyoto University, Kyoto 606-8502, Japan\\
      ${}^b$ Department of Physics, Shinshu University,\\
      Matsumoto 390-8621, Japan
    \end{center}}
\newcommand{\Accepted}[1]{\begin{center}
    {\large \sf #1}\\ \vspace{1mm}{\small \sf Accepted for Publication}
    \end{center}}

\preprint
\thispagestyle{empty}

\Title{Modification of Crum's Theorem for `Discrete' Quantum Mechanics}
\Author

\Address
\vspace{1cm}

\begin{abstract}
Crum's theorem in one-dimensional quantum mechanics asserts the existence
of an associated Hamiltonian system for any given Hamiltonian with the
complete set of eigenvalues and eigenfunctions.
The associated system is iso-spectral to the original one except for the
lowest energy state, which is deleted. A modification due to Krein-Adler
provides algebraic construction of a new complete Hamiltonian system by
deleting a finite number of energy levels.
Here we present a discrete version of the modification based on the
Crum's theorem for the `discrete' quantum mechanics developed by two of
the present authors.
\end{abstract}

\section{Introduction}
\label{intro}

Crum's seminal paper of 1955 \cite{crum} has played an essential role
in elucidating the structure of one-dimensional quantum mechanical
systems in general and exactly solvable ones, in particular.
Throughout this paper, we mean `exact solvability' in the Schr\"odinger
picture, namely a quantum system is exactly solved when the complete
set of eigenvalues and eigenfunctions are known.
Many exactly solvable quantum mechanical Hamiltonians were constructed
and investigated by combining shape invariance \cite{genden} and Crum's
theorem \cite{crum,Darboux}, or the factorisation method \cite{infhul}
or the method of the so-called supersymmetric quantum mechanics
\cite{susyqm}. It is interesting to note that most of these shape
invariant systems are also solvable in the Heisenberg picture \cite{os7}.
Exactly solvable quantum mechanical systems of one and many degrees
of freedom are not only important in their own right but also have
fundamental applications in various disciplines of physics/mathematics,
{\em e.g.} the Fokker-Planck equations \cite{risken} and their
discretised version, birth and  death processes \cite{bdproc},
to name a few.

Shape invariance is a sufficient condition for exactly solvable
quantum mechanical systems. The number of shape invariant systems,
however, was quite limited; only about a dozen until the recent discovery
\cite{gomez,quesne,quesne2} of the several types of infinitely many
shape invariant Hamiltonians \cite{os16,os17,os18,os19,hos} which
led to the infinitely many exceptional Laguerre, Jacobi, Wilson and
Askey-Wilson polynomials.
Many methods were proposed to derive exactly solvable but non-shape
invariant quantum mechanical systems from known shape invariant ones
\cite{dubov,adler,nieto,bagsam,junkroy}. (We apologise to those
whose work we have missed.)
Among them Krein-Adler's modification \cite{adler} of Crum's theorem
is the most comprehensive way to generate infinitely many variants of
exactly solvable Hamiltonians and their eigenfunctions, starting from
an exactly solvable one.
The derived system is {\em iso-spectral} with the original one except
that a finite number of energy levels are deleted. If the original
system has polynomial eigenfunctions, as is usually the case, the
derived systems have also polynomial eigenfunctions. By construction,
these polynomials constitute a complete set of orthogonal functions.
But they do not qualify to be called {\em exceptional orthogonal
polynomials\/} \cite{os16,os17,os18,os19,hos} since some members of
certain degrees are  {\em missing} due to the {\em deletion\/}.

The discrete quantum mechanics is a deformation of the ordinary
quantum mechanics in the sense that the Schr\"odinger equation is
a second order {\em difference\/} equation instead of differential.
In the formulation of Odake and Sasaki \cite{os4,os6,os12,os13},
the algebraic and analytical structure of quantum mechanics as well
as shape invariance and exact solvability are retained in the discrete
version. The eigenfunctions of the exactly solvable one-dimensional
discrete quantum mechanics are the Askey-scheme of hypergeometric
orthogonal polynomials and their $q$-versions \cite{nikiforov,askey,
ismail,koeswart}, {\em e.g.\/} the continuous Hahn, the Wilson and
the Askey-Wilson polynomials.
These examples are all shape invariant and they are also solvable in
the Heisenberg picture \cite{os4,os7}. The dynamical symmetry algebra
of these algebras are the Askey-Wilson algebras \cite{zhedanov,os14}
and degenerations, which contain the $q$-oscillator algebra \cite{os11}.
The discrete version of Crum's theorem was also established recently
\cite{matveev,os15}.

\bigskip
In this paper we present the discrete quantum mechanics version of
Adler's modification \cite{adler} of Crum's theorem. It allows to
generate an infinite variety of exactly solvable discrete quantum
Hamiltonian systems. The insight obtained from Crum's theorems and
their modification, in the ordinary and the discrete quantum mechanics,
is essential for the recent derivation of the infinite numbers of
shape invariant systems and the new exceptional orthogonal polynomials
\cite{os16,os17,os19}. We will discuss the main results, the
specialisation to the cases of polynomial eigenfunctions and simplest
example for various exactly solvable cases; first for the ordinary
quantum mechanics and then for the discrete versions.
The reason is two-fold; firstly to introduce appropriate notion and
notation in the familiar cases of the ordinary quantum mechanics.
Secondly we choose to reveal the underlying logical processes which
are not easy to fathom in Adler's paper \cite{adler} or in Crum's
original article \cite{crum}. As seen in the subsequent sections,
the logical structures of the associated Hamiltonian systems and
their modification by {\em deletion\/} of energy levels are shared
by the ordinary and the discrete quantum mechanics.

This paper is organised as follows. In section two, Adler's modification
of Crum's theorem is recapitulated in appropriate notation for our
purposes. The specialisation to the cases of polynomial eigenfunctions
is discussed in some detail.
Section three provides the discrete quantum mechanics version of
the modification of Crum's theorem. Again the specialisation to the
cases of polynomial eigenfunctions is mentioned.
Appendix gives the simplest examples of the modified Hamiltonian systems
obtained by deleting the lowest lying $\ell$ excited states for various
exactly solvable Hamiltonians.
Appendix A provides three examples from the ordinary quantum mechanics,
the harmonic oscillator, the radial oscillator, the
Darboux-P\"oschl-Teller potential.
Appendix B is for the four examples from the discrete quantum mechanics,
the Hamiltonians of the Meixner-Pollaczek, the continuous Hahn, the Wilson
and the Askey-Wilson polynomials \cite{os4,os13}, which are known to reduce
to the Hermite, the Laguerre and the Jacobi polynomials in certain
limits, respectively.

\section{Ordinary Quantum Mechanics}
\label{ordQM}
\setcounter{equation}{0}

\subsection{Adler's modification of Crum's theorem}
\label{genordQM}

Let us start with a generic one-dimensional quantum mechanical (QM)
system having discrete semi-infinite energy levels only:
\begin{equation}
  0=\mathcal{E}_0 <\mathcal{E}_1 < \mathcal{E}_2 < \cdots.
  \label{semipositive}
\end{equation}
Here we have chosen the constant part of the Hamiltonian so that
the groundstate energy is zero. Then the Hamiltonian is {\em positive
semi-definite\/} and can be factorised,
\begin{align}
  \mathcal{H}&=p^2+U(x)
  =p^2+\Bigl(\frac{d\mathcal{W}(x)}{dx}\Bigr)^2
  +\frac{d^2\mathcal{W}(x)}{dx^2},\qquad p=-i\frac{d}{dx},\\
  &=\mathcal{A}^{\dagger}\mathcal{A},\qquad\quad
  \mathcal{A}\eqdef\frac{d}{dx}-\frac{d\mathcal{W}(x)}{dx},\quad
  \mathcal{A}^{\dagger}=-\frac{d}{dx}-\frac{d\mathcal{W}(x)}{dx}.
\end{align}
Here a real and smooth function $\mathcal{W}(x)\in\mathbb{C}^\infty$ is
called a {\em prepotential\/} and it parametrises the groundstate
wavefunction $\phi_0(x)$, which has {\em no node\/} and can be chosen
real and positive:
\begin{equation}
  \phi_0(x)=e^{\mathcal{W}(x)}.
  \label{prepotdef}
\end{equation}
It is trivial to verify
\begin{equation}
  \mathcal{A}\phi_0(x)=0\ \Rightarrow\ \mathcal{H}\phi_0(x)=0.
\end{equation}
In one dimension all the energy levels are non-degenerate.
By construction all the eigenfunctions are square-integrable and
orthogonal with each other and form a complete basis of the Hilbert space:
\begin{alignat}{2}
  \mathcal{H}\phi_n(x)&=\mathcal{E}_n\phi_n(x),
  &\quad n&\in\mathbb{Z}_+,\\
  \int_{x_1}^{x_2}\!\!\phi_n(x)^*\phi_m(x)dx&=h_n\delta_{nm},\quad
  0<h_n<\infty,&\quad n,m&\in\mathbb{Z}_+,
\end{alignat}
where $\mathbb{Z}_+$ is the set of non-negative integers $\{0,1,2,\ldots\}$.
It is well-known that the $n$-th excited wavefunction $\phi_n(x)$ has
$n$ nodes in the interior.
For simplicity we choose all the eigenfunctions to be real.
A few exactly solvable examples are given in Appendix.

\bigskip
Let us choose a set of $\ell$ distinct non-negative integers
$\mathcal{D}\eqdef\{d_1,d_2,\ldots,d_\ell\}\subset \mathbb{Z}_+$,
satisfying the condition
\begin{equation}
  \prod_{j=1}^\ell(m-d_j)\ge0,\quad\forall m\in\mathbb{Z}_+.
  \label{dellcond}
\end{equation}
This condition means that the set $\mathcal{D}$ consists of several
clusters, each containing an {\em even number\/} of {\em contiguous\/}
integers
\begin{equation}
  d_{k_1}, d_{k_1}+1,\cdots,d_{k_2}\ ;\ d_{k_3},
  d_{k_3}+1,\cdots, d_{k_4}\ ;\ d_{k_5},
  d_{k_5}+1,\cdots, d_{k_6},\ ;\ \cdots,
\end{equation}
where $d_{k_2}+1<d_{k_3},\ d_{k_4}+1<d_{k_5},\ \cdots$.
If $d_{k_1}=0$ for the lowest lying cluster, it could contain an even
or odd number of contiguous integers.
The set $\mathcal{D}$ specifies the energy levels to be {\em deleted}.
This simply reflects the fact that no singularity arises when two
neighbouring levels are deleted. In the ordinary QM, the zeros of the
two neighbouring eigenfunctions $\phi_{j}$ and $\phi_{j+1}$ interlace
with each other. This fact is essential for the non-singularity of the
potential after deletion. See Adler's paper \cite{adler} for a proof.
The situation is essentially the same in the discrete QM. However, in
the dQM to be discussed in the subsequent section, due to the lack of
general theorem, the interlacing of the zeros of the two neighbouring
eigenfunctions $\phi_{j}$ and $\phi_{j+1}$ must be verified for each
specific Hamiltonian.
Deleting an arbitrary number of contiguous energy levels
starting from the groundstate ($d_{k_1}=0$) is achieved by the
original Crum's theorem \cite{crum}.

Next we will construct Hamiltonian systems corresponding to the
successive deletions $\mathcal{H}_{d_1\ldots}$ (and
$\mathcal{A}_{d_1\ldots}$, $\mathcal{A}_{d_1\ldots}^{\dagger}$, etc.)
step by step, algebraically. It should be noted that some quantities
in the intermediate steps could be singular.
For given $d_1$  the first Hamiltonian $\mathcal{H}$ can be expressed
in two different ways:
\begin{align}
  &\mathcal{H}=\mathcal{A}^{\dagger}\mathcal{A}
  =\mathcal{A}^{\dagger}_{d_1}\mathcal{A}_{d_1}
  +\mathcal{E}_{d_1},\quad\mathcal{A}_{d_1}\phi_{d_1}=0,
  \label{H=Ad1dAd1}\\
  &\mathcal{A}_{d_1}\eqdef\frac{d}{dx}-\frac{d\mathcal{W}_{d_1}(x)}{dx},
  \quad\mathcal{A}_{d_1}^{\dagger}\eqdef-\frac{d}{dx}
  -\frac{d\mathcal{W}_{d_1}(x)}{dx},\quad
  \mathcal{W}_{d_1}(x)\eqdef\log\phi_{d_1}(x),
  \label{Ad1Ad1d}\\
  &U(x)=\Bigl(\frac{d\mathcal{W}(x)}{dx}\Bigr)^2
  +\frac{d^2\mathcal{W}(x)}{dx^2}
  =\Bigl(\frac{d\mathcal{W}_{d_1}(x)}{dx}\Bigr)^2
  +\frac{d^2\mathcal{W}_{d_1}(x)}{dx^2}+\mathcal{E}_{d_1}.
\end{align}
Unless $d_1=0$, $\mathcal{W}_{d_1}(x)$ is singular due to the zeros
of $\phi_{d_1}(x)$.
It is very important to note that $\mathcal{A}_{d_1}^{\dagger}$
in \eqref{Ad1Ad1d} is a `formal adjoint' of $\mathcal{A}_{d_1}$.
We stick to this notation, since the algebraic structure of various
expressions appearing in the deletion processes, from \eqref{H=Ad1dAd1} to
\eqref{fQMHb1bnphi=..}, are best described by using the `formal adjoint'.
These define a new Hamiltonian system
\begin{align}
  &\mathcal{H}_{d_1}
  \eqdef \mathcal{A}_{d_1}\mathcal{A}_{d_1}^{\dagger}
  +\mathcal{E}_{d_1}=p^2+U_{d_1}(x),\\
  &U_{d_1}(x) \eqdef\Bigl(\frac{d\mathcal{W}_{d_1}(x)}{dx}\Bigr)^2
  -\frac{d^2\mathcal{W}_{d_1}(x)}{dx^2}+\mathcal{E}_{d_1},
\end{align}
with the `eigenfunctions'
\begin{equation}
  \mathcal{H}_{d_1}\phi_{d_1\, n}(x)=\mathcal{E}_n\phi_{d_1\, n}(x),
  \quad\phi_{d_1\, n}(x)\eqdef\mathcal{A}_{d_1}\phi_n(x),\quad
  n\in\mathbb{Z}_+\backslash\{d_1\}.
\end{equation}
Note that the energy level $d_1$ is now deleted,
$\phi_{d_1\, d_1}(x)\equiv0$,  from the set of `eigenfunctions'
$\{\phi_{d_1\, n}(x)\}$ of the new Hamiltonian $\mathcal{H}_{d_1}$.

Suppose we have determined  $\mathcal{H}_{d_1\,\ldots\,d_s}$ and
$\phi_{d_1\,\ldots\,d_s\,n}(x)$ with $s$ deletions. They have the
following properties
\begin{align}
  &\mathcal{H}_{d_1\,\ldots\,d_s}
  \eqdef\mathcal{A}_{d_1\,\ldots\,d_s}
  \mathcal{A}_{d_1\,\ldots\,d_s}^{\dagger}+\mathcal{E}_{d_s}
  \eqdef p^2+U_{d_1\,\ldots\,d_s}(x),
  \label{QMHb1bsdef}\\
  &\mathcal{W}_{d_1\,\ldots\,d_s}(x)\eqdef\log\phi_{d_1\,\ldots\,d_{s}}(x),\\
  &\mathcal{A}_{d_1\,\ldots\,d_s}\eqdef\frac{d}{dx}
  -\frac{d\mathcal{W}_{d_1\,\ldots\,d_s}(x)}{dx},
  \quad\mathcal{A}_{d_1\,\ldots\,d_s}^{\dagger}\eqdef-\frac{d}{dx}
  -\frac{d\mathcal{W}_{d_1\,\ldots\,d_s}(x)}{dx},\\
  &U_{d_1\,\ldots\,d_s}(x)
  \eqdef\Bigl(\frac{d\mathcal{W}_{d_1\,\ldots\,d_s}(x)}{dx}\Bigr)^2
  -\frac{d^2\mathcal{W}_{d_1\,\ldots\,d_s}(x)}{dx^2}+\mathcal{E}_{d_s},\\
  &\phi_{d_1\,\ldots\,d_s\,n}(x)
  \eqdef\mathcal{A}_{d_1\,\ldots\,d_s}\phi_{d_1\,\ldots\,d_{s-1}\,n}(x)
  \quad(n\in \mathbb{Z}_+\backslash\{d_1,\ldots,d_s\}),
  \label{QMphib1bsndef}\\
  &\mathcal{H}_{d_1\,\ldots\,d_s}\phi_{d_1\,\ldots\,d_s\,n}(x)
  =\mathcal{E}_n\phi_{d_1\,\ldots\,d_s\,n}(x)
  \quad(n\in \mathbb{Z}_+\backslash\{d_1,\ldots,d_s\}).
  \label{QMHb1bnphi=..}
\end{align}
We have also
\begin{equation}
  \phi_{d_1\,\ldots\,d_{s-1}\,n}(x)
  =\frac{\mathcal{A}_{d_1\,\ldots\,d_s}^{\dagger}}
  {\mathcal{E}_n-\mathcal{E}_{d_s}}\phi_{d_1\,\ldots\,d_s\,n}(x)
  \quad(n\in \mathbb{Z}_+\backslash\{d_1,\ldots,d_s\}).
\end{equation}
Next we will define a new Hamiltonian system with one more deletion of
the level $d_{s+1}$. We can show
\begin{align}
  &\mathcal{H}_{d_1\,\ldots\,d_s}
  =\mathcal{A}_{d_1\,\ldots\,d_s\,d_{s+1}}^{\dagger}
  \mathcal{A}_{d_1\,\ldots\,d_s\,d_{s+1}}
  +\mathcal{E}_{d_{s+1}},\quad
  \mathcal{A}_{d_1\,\ldots\,d_s\,d_{s+1}}\phi_{d_1\,\ldots\,d_s\,d_{s+1}}=0,
  \label{QMHb1bs}\\
  &\mathcal{A}_{d_1\,\ldots\,d_s\,d_{s+1}}\eqdef\frac{d}{dx}
  -\frac{d\mathcal{W}_{d_1\,\ldots\,d_s\,d_{s+1}}(x)}{dx},\quad
  \mathcal{A}_{d_1\,\ldots\,d_s\,d_{s+1}}^{\dagger}\eqdef-\frac{d}{dx}
  -\frac{d\mathcal{W}_{d_1\,\ldots\,d_s\,d_{s+1}}(x)}{dx},\\
  &\mathcal{W}_{d_1\,\ldots\,d_s\,d_{s+1}}(x)\eqdef
  \log\phi_{d_1\,\ldots\,d_s\,d_{s+1}}(x),
  \label{Wb1bs+1}\\
  &U_{d_1\,\ldots\,d_s}(x)
  =\Bigl(\frac{d\mathcal{W}_{d_1\,\ldots\,d_s\,d_{s+1}}(x)}{dx}\Bigr)^2
  +\frac{d^2\mathcal{W}_{d_1\,\ldots\,d_s\,d_{s+1}}(x)}{dx^2}
  +\mathcal{E}_{d_{s+1}}.
\end{align}
These determine a new Hamiltonian system with $s+1$ deletions:
\begin{align}
  &\mathcal{H}_{d_1\,\ldots\,d_{s+1}}
  \eqdef\mathcal{A}_{d_1\,\ldots\,d_{s+1}}
  \mathcal{A}_{d_1\,\ldots\,d_{s+1}}^{\dagger}+\mathcal{E}_{d_{s+1}}
  \eqdef p^2+U_{d_1\,\ldots\,d_{s+1}}(x),
  \label{2QMHb1bsdef}\\
  &U_{d_1\,\ldots\,d_{s+1}}(x)
  \eqdef\Bigl(\frac{d\mathcal{W}_{d_1\,\ldots\,d_{s+1}}(x)}{dx}\Bigr)^2
  -\frac{d^2\mathcal{W}_{d_1\,\ldots\,d_{s+1}}(x)}{dx^2}
  +\mathcal{E}_{d_{s+1}},\\
  &\phi_{d_1\,\ldots\,d_{s+1}\,n}(x)
  \eqdef\mathcal{A}_{d_1\,\ldots\,d_{s+1}}\phi_{d_1\,\ldots\,d_{s}\,n}(x)
  \quad(n\in \mathbb{Z}_+\backslash\{d_1,\ldots,d_{s+1}\}),
  \label{2QMphib1bsndef}\\
  &\mathcal{H}_{d_1\,\ldots\,d_{s+1}}\phi_{d_1\,\ldots\,d_{s+1}\,n}(x)
  =\mathcal{E}_n\phi_{d_1\,\ldots\,d_{s+1}\,n}(x)
  \quad(n\in \mathbb{Z}_+\backslash\{d_1,\ldots,d_{s+1}\}).
  \label{2QMHb1bnphi=..}
\end{align}

\bigskip
After deleting all the $\mathcal{D}=\{d_1,\,\cdots,\,d_\ell\}$ energy
levels, the resulting Hamiltonian system
$\mathcal{H}_{\mathcal{D}}\equiv\mathcal{H}_{d_1\,\ldots\,d_{\ell}}$,
$\mathcal{A}_{\mathcal{D}}\equiv\mathcal{A}_{d_1\,\ldots\,d_{\ell}}$,
etc has the following form:
\begin{align}
  &\mathcal{H}_{\mathcal{D}}
  \eqdef\mathcal{A}_{\mathcal{D}}
  \mathcal{A}_{\mathcal{D}}^{\dagger}+\mathcal{E}_{d_{\ell}}
  \eqdef p^2+U_{\mathcal{D}}(x),
  \label{fQMHb1bsdef}\\
  &U_{\mathcal{D}}(x)
  \eqdef\Bigl(\frac{d\mathcal{W}_{\mathcal{D}}(x)}{dx}\Bigr)^2
  -\frac{d^2\mathcal{W}_{\mathcal{D}}(x)}{dx^2}+\mathcal{E}_{d_{\ell}},\quad
  \mathcal{W}_{\mathcal{D}}(x)\eqdef\log\phi_{d_1\,\cdots\,d_\ell}(x),\\
  &\phi_{{\mathcal{D}}\,n}(x)
  \eqdef\mathcal{A}_{\mathcal{D}}\phi_{d_1\,\cdots\,d_{\ell-1}\,n}(x)
  \quad(n\in\mathbb{Z}_+\backslash\mathcal{D}),
  \label{fQMphib1bsndef}\\
  &\mathcal{H}_{\mathcal{D}}\phi_{{\mathcal{D}}\,n}(x)
  =\mathcal{E}_n\phi_{{\mathcal{D}}\,n}(x)
  \quad(n\in\mathbb{Z}_+\backslash\mathcal{D}).
  \label{fQMHb1bnphi=..}
\end{align}
Now that $\mathcal{H}_{\mathcal{D}}$ has the lowest energy level $\mu$:
\begin{equation}
  \mu\eqdef\min\{n\,|\,n\in\mathbb{Z}_+\backslash\mathcal{D}\},
\end{equation}
with the groundstate wavefunction $\bar{\phi}_\mu(x)$
\begin{equation}
  \bar{\phi}_\mu(x)\eqdef\phi_{\mathcal{D}\,\mu}(x)\equiv
  \phi_{d_1\,\cdots\,d_\ell\, \mu}(x).
\end{equation}
As usual the Hamiltonian system can be expressed simply in terms of
the groundstate wavefunction $\bar{\phi}_\mu(x)$, which we will denote
by new symbols $\bar{\mathcal H}$, $\bar{\mathcal A}$, etc:
\begin{align}
  &\bar{\mathcal{H}}\equiv\mathcal{H}_{\mathcal{D}}
  \eqdef\bar{\mathcal{A}}^\dagger
  \bar{\mathcal{A}}+\mathcal{E}_{\mu}
  \eqdef p^2+\bar{U}(x),
  \label{bfQMHb1bsdef}\\
  &\bar{\mathcal{A}}\equiv\mathcal{A}_{\mathcal{D}\,\mu}
  \eqdef\frac{d}{dx}-\frac{d\bar{\mathcal{W}}(x)}{dx},\quad
  \bar{\mathcal{A}}^{\dagger}\equiv\mathcal{A}_{\mathcal{D}\,\mu}^{\dagger}
  \eqdef-\frac{d}{dx}-\frac{d\bar{\mathcal{W}}(x)}{dx},\\
  &\bar{U}(x)\equiv U_{\mathcal{D}\,\mu}(x)
  \eqdef\Bigl(\frac{d\bar{\mathcal{W}}(x)}{dx}\Bigr)^2
  +\frac{d^2\bar{\mathcal{W}}(x)}{dx^2}+\mathcal{E}_{\mu},\quad
  \bar{\mathcal{W}}(x)\equiv\mathcal{W}_{\mathcal{D}\,\mu}(x)
  \eqdef\log\bar{\phi}_\mu(x),\\
  &\bar{\mathcal{H}}\bar{\phi}_{n}(x)
  =\mathcal{E}_n\bar{\phi}_{n}(x),\quad
  \bar{\phi}_{n}(x)\equiv\phi_{{\mathcal{D}}\,n}(x)
  \qquad(n\in \mathbb{Z}_+\backslash\mathcal{D}).
  \label{bfQMHb1bnphi=..}
\end{align}

As shown by Krein-Adler \cite{adler}, the results can be expressed
succinctly:
\begin{align}
  &\bar{\phi}_{n}(x)
  =\frac{\text{W}[\phi_{d_1},\phi_{d_2},\ldots,\phi_{d_\ell},\phi_n](x)}
  {\text{W}[\phi_{d_1},\phi_{d_2},\ldots,\phi_{d_\ell}](x)}
  \quad(n\in\mathbb{Z}_+\backslash\mathcal{D}),
  \label{QMphib1bsn}\\
  &\bar{U}(x)\equiv U_{d_1,\ldots,d_\ell}(x)=U(x)
  -2\frac{d^2}{dx^2}\Bigl(
  \log\text{W}\,[\phi_{d_1},\phi_{d_2},\ldots,\phi_{d_\ell}](x)\Bigr)
  \quad(\ell\geq 0),
  \label{Ub1bs}
\end{align}
in which the Wronskian determinant is defined by
\begin{equation}
  \text{W}\,[f_1,\ldots,f_n](x)
  \eqdef\det\Bigl(\frac{d^{j-1}f_k(x)}{dx^{j-1}}\Bigr)_{1\leq j,k\leq n}.
\end{equation}
For $n=0$, we set $\text{W}\,[\cdot](x)=1$.
In deriving the determinant formulas \eqref{QMphib1bsn} and \eqref{Ub1bs}
use is made of the properties of the Wronskian
\begin{align}
  &\text{W}[gf_1,gf_2,\ldots,gf_n](x)
  =g(x)^n\text{W}[f_1,f_2,\ldots,f_n](x),
  \label{Wformula1}\\
  &\text{W}\bigl[\text{W}[f_1,f_2,\ldots,f_n,g],
  \text{W}[f_1,f_2,\ldots,f_n,h]\,\bigr](x)\n
  &=\text{W}[f_1,f_2,\ldots,f_n](x)\,
  \text{W}[f_1,f_2,\ldots,f_n,g,h](x)
  \qquad(n\geq 0).
  \label{Wformula2}
\end{align}
Let us note that $U_{d_1\,\ldots\,d_\ell}(x)$ and
$\phi_{d_1,\ldots\,d_\ell\,n}(x)$ are symmetric with respect to
$d_1,\ldots,d_\ell$, and thus
$\bar{\mathcal H}\equiv\mathcal{H}_{d_1\,\ldots\,d_\ell}$ is independent
of the order of $\{d_j\}$.

Let us state Adler's theorem again;
{\it
If the set of deleted energy levels $d_1,\ldots,d_\ell$ satisfy
the condition \eqref{dellcond}, the Hamiltonian
$\bar{\mathcal H}\equiv \mathcal{H}_{d_1\,\ldots\,d_\ell}=p^2+\bar{U}(x)$
with \eqref{Ub1bs} is well-defined and hermitian, and its complete set
of eigenfunctions \eqref{bfQMHb1bnphi=..} are given by \eqref{QMphib1bsn}.
}
Crum's theorem corresponds to the choice
$\{d_1,\ldots,d_\ell\}=\{0,1,\ldots,\ell-1\}$, and the resulting lowest
energy level is $\mu=\ell$ and there is no deleted energy levels above
the new groundstate.

\subsection{Polynomial eigenfunctions}

In this subsection we consider the typical case of shape invariant systems
in which the eigenfunctions consist of the orthogonal polynomials $\{P_n\}$:
\begin{equation}
  \phi_n(x)=\phi_0(x)P_n(\eta(x)),\quad\phi_0(x)=e^{\mathcal{W}(x)},
\end{equation}
in which $\eta(x)$ is called the {\em sinusoidal coordinate\/}.
As shown in detail in the examples in Appendix A, $\eta(x)=x$ for the
harmonic oscillator (the Hermite polynomials), $\eta(x)=x^2$ for the
radial oscillator (the Laguerre polynomials) and $\eta(x)=\cos2x$ for
the Darboux-P\"oschl-Teller potential (the Jacobi polynomials).
The groundstate wavefunction $\phi_0(x)$ provides the orthogonality
weight function
\begin{equation}
  \int_{x_1}^{x_2}\!\!\phi_{0}(x)^2\,P_n(\eta(x))P_m(\eta(x))dx
  =h_n\delta_{nm},\quad n,\,m\in\mathbb{Z}_+.
\end{equation}

In this case, the modification of Crum's theorem produces the
eigenfunctions $\{\bar{\phi}_n(x)\}$ which again consist of polynomials
in $\eta(x)$. By using \eqref{Wformula1} and
\begin{equation}
  \check{f}_j(x)\eqdef f_j(\eta(x)),
  \ \ \text{W}[\check{f}_1,\check{f}_2,\ldots,\check{f}_n](x)
  =\bigl(\tfrac{d\eta(x)}{dx}\bigr)^{\frac12n(n-1)}
  \text{W}[f_1,f_2,\ldots,f_n](\eta(x)),
  \label{Wformula3}
\end{equation}
we obtain a simple expression of the eigenfunctions
\begin{equation}
  \bar{\phi}_n(x)=\phi_0(x)\bigl(\tfrac{d\eta(x)}{dx}\bigr)^\ell
  \frac{\text{W}[P_{d_1},P_{d_2},\ldots,P_{d_\ell},P_n](\eta(x))}
  {\text{W}[P_{d_1},P_{d_2},\ldots,P_{d_\ell}](\eta(x))}.
\end{equation}
This simply means that the resulting eigenfunctions are again polynomials
in $\eta(x)$:
\begin{align}
  \bar{\phi}_n(x)&=\bar{\psi}(x)\mathcal{P}_n(\eta(x)),\\
  \bar{\psi}(x)&\eqdef\frac{\phi_0(x)\bigl(\tfrac{d\eta(x)}{dx}\bigr)^\ell}
  {\text{W}[P_{d_1},P_{d_2},\ldots,P_{d_\ell}](\eta(x))},\quad
  \mathcal{P}_n(\eta)\eqdef
  \text{W}[P_{d_1},P_{d_2},\ldots,P_{d_\ell},P_n](\eta),
  \label{polyphi0}
\end{align}
satisfying the orthogonality relation
\begin{equation}
  \int_{x_1}^{x_2}\bar{\psi}(x)^2\,
  \mathcal{P}_n(\eta(x))\mathcal{P}_m(\eta(x))dx
  =\bar{h}_n\delta_{nm},\quad n,\,m\in\mathbb{Z}_+\backslash\mathcal{D}.
\end{equation}
Let us emphasise that $n$ is not the {\em degree} in $\eta$ and by
construction, $\ell$ members are missing:
$\mathcal{P}_{d_1}=\mathcal{P}_{d_2}=\cdots=\mathcal{P}_{d_\ell}\equiv0$.
Therefore these polynomials cannot be called {\em exceptional orthogonal
polynomials} \cite{gomez,os16,os19}.

\section{`Discrete' Quantum Mechanics}
\label{discrQM}
\setcounter{equation}{0}

Let us begin with a few general remarks on the one-dimensional discrete
QM with pure imaginary shifts. See \cite{os13} for the general
introduction to the discrete quantum mechanics with pure imaginary
shifts and \cite{os15} for the Crum's theorem in the discrete QM.
In the discrete QM, the dynamical variables are, as in the ordinary QM,
the coordinate $x$, which takes value in an infinite or a semi-infinite
or a finite range of the real axis and the canonical momentum $p$, which
is realised as a differential operator $p=-i\partial_x$.
Since the momentum operator appears in exponentiated forms
$e^{\pm \gamma p}$, $\gamma\in\mathbb{R}$, in a Hamiltonian, it causes
finite pure imaginary shifts in the wavefunction
$e^{\pm \gamma p}\psi(x)=\psi(x\mp i\gamma)$.
This requires the wavefunction as well as other functions appearing
in the Hamiltonian to be {\em analytic\/} in $x$ within a certain
complex domain including the physical region of the coordinate.
Let us introduce the $*$-operation on an analytic function,
$*:f\mapsto f^*$. If $f(x)=\sum\limits_{n}a_nx^n$, $a_n\in\mathbb{C}$,
then $f^*(x)\eqdef\sum\limits_{n}a_n^*x^n$, in which $a_n^*$ is the complex
conjugation of $a_n$. Obviously $f^{**}(x)=f(x)$ and $f(x)^*=f^*(x^*)$.
If a function satisfies $f^*=f$, we call it a `real' function, for it
takes real values on the real line.

The starting point is again a generic one dimensional discrete quantum
mechanical Hamiltonian with discrete semi-infinite energy levels only
\eqref{semipositive}. Again we assume that the groundstate energy is
chosen to be zero $\mathcal{E}_0=0$, so that the Hamiltonian is positive
semi-definite. The generic factorised Hamiltonian reads \cite{os4,os13}
\begin{align}
  &\mathcal{H}=\mathcal{A}^{\dagger}\mathcal{A}
  =\sqrt{V(x)}\,e^{\gamma p}\sqrt{V^*(x)}
  +\!\sqrt{V^*(x)}\,e^{-\gamma p}\sqrt{V(x)}-V(x)-V^*(x),
  \label{discrham}\\
  &\mathcal{A}\eqdef i\bigl(e^{\frac{\gamma}{2}p}\sqrt{V^*(x)}
  -e^{-\frac{\gamma}{2}p}\sqrt{V(x)}\,\bigr),\quad
  \mathcal{A}^{\dagger}\eqdef -i\bigl(\sqrt{V(x)}\,e^{\frac{\gamma}{2}p}
  -\sqrt{V^*(x)}\,e^{-\frac{\gamma}{2}p}\bigr).
\end{align}
Since the $*$-operation for $\mathcal{A}f$, $\mathcal{A}^{\dagger}f$ and
$\mathcal{H}f$ satisfies
\begin{equation}
  (\mathcal{A}f)^*(x)=\mathcal{A}f^*(x),\quad
  (\mathcal{A}^{\dagger}f)^*(x)=\mathcal{A}^{\dagger}f^*(x),\quad
  (\mathcal{H}f)^*(x)=\mathcal{H}f^*(x),
\end{equation}
they map a `real' function to a `real' function
\begin{equation}
  f^*=f\quad\Rightarrow\quad
  (\mathcal{A}f)^*=\mathcal{A}f,
  \ \ (\mathcal{A}^{\dagger}f)^*=\mathcal{A}^{\dagger}f,
  \ \ (\mathcal{H}f)^*=\mathcal{H}f.
  \label{f*=f==>}
\end{equation}
By specifying the function $V(x)$, various explicit examples are
obtained \cite{os4,os13,os7}. 
A few exactly solvable examples are given in Appendix.
The corresponding Schr\"{o}dinger equation
$\mathcal{H}\psi(x)=\mathcal{E}\psi(x)$ is a {\em difference equation\/}
\begin{align}
  \sqrt{V(x)V^*(x-i\gamma)}\,\psi(x-i\gamma)
  &+\sqrt{V^*(x)V(x+i\gamma)}\,\psi(x+i\gamma)\n
  &-\bigl(V(x)+V^*(x)\bigr)\psi(x)=\mathcal{E}\psi(x),
  \label{difschro}
\end{align}
instead of differential in the ordinary QM.
Although this equation looks rather complicated, the equation for the
polynomial eigenfunctions \eqref{polyeq} has a familiar form of
difference equations.
Again the groundstate wavefunction $\phi_0(x)$ is determined as a
zero mode of $\mathcal{A}$, $\mathcal{A}\phi_0(x)=0$ ($\Rightarrow$
$\mathcal{H}\phi_0(x)=0$), namely,
\begin{equation}
  \sqrt{V^*(x-i\tfrac{\gamma}{2})}\phi_0(x-i\tfrac{\gamma}{2})-
  \sqrt{V(x+i\tfrac{\gamma}{2})}\phi_0(x+i\tfrac{\gamma}{2})=0.
  \label{explizero}
\end{equation}
This dictates how the `phase' of the potential function $V$ is related
to that of the groundstate wavefunction $\phi_0$.
Here we also assume that the groundstate wavefunction $\phi_0(x)$
has no node and chosen to be real and positive for real $x$.

Due to the lack of generic theorems in the theory of difference equations,
let us assume that all the energy levels are non-degenerate and that all
the eigenfunctions are square-integrable and orthogonal with each other
and form a complete basis of the Hilbert space:
\begin{alignat}{2}
  \mathcal{H}\phi_n(x)&=\mathcal{E}_n\phi_n(x),
  &\quad n&\in \mathbb{Z}_+,\\
  \int_{x_1}^{x_2}\!\!\phi_n(x)^*\phi_m(x)dx&=h_n\delta_{nm},\quad
  0<h_n<\infty,&\quad n,m&\in \mathbb{Z}_+.
\end{alignat}
In most explicit examples these statements can be verified
straightforwardly. For simplicity we choose all the eigenfunctions to be
real on the real axis $\phi^*_n=\phi_n$, which is made possible by
\eqref{f*=f==>}.

\subsection{Modification of Crum's theorem}
\label{gendQM}

The formulation of the modified Crum's theorem in the discrete quantum
mechanics goes almost parallel to that in the ordinary quantum mechanics.
Again the presentation is purely algebraic. Let us note that various
quantities in intermediate steps might have singularities and
Hamiltonians might not be hermitian. We choose a set of distinct
non-negative integers
$\mathcal{D}\eqdef\{d_1,d_2,\ldots,d_\ell\}\subset\mathbb{Z}_+$,
satisfying the condition \eqref{dellcond} as before.
First let us note that the Hamiltonian  $\mathcal{H}$ can be rewritten
by incorporating the level $d_1$ as:
\begin{align}
  &\mathcal{H}=\mathcal{A}_{d_1}^{\dagger}\mathcal{A}_{d_1}
  +\mathcal{E}_{d_1},\quad\mathcal{A}_{d_1}\phi_{d_1}=0,\\
  &\mathcal{A}_{d_1}\eqdef i\bigl(e^{\frac{\gamma}{2}p}\sqrt{V_{d_1}^*(x)}
  -e^{-\frac{\gamma}{2}p}\sqrt{V_{d_1}(x)}\,\bigr),\quad
  \mathcal{A}_{d_1}^{\dagger}\eqdef
  -i\bigl(\sqrt{V_{d_1}(x)}\,e^{\frac{\gamma}{2}p}
  -\sqrt{V_{d_1}^*(x)}\,e^{-\frac{\gamma}{2}p}\bigr),\\
  &V_{d_1}(x)\eqdef\sqrt{V(x)V^*(x-i\gamma)}\,
  \frac{\phi_{d_1}(x-i\gamma)}{\phi_{d_1}(x)}.
\end{align}
These define a new Hamiltonian system
\begin{align}
  &\mathcal{H}_{d_1}
  \eqdef\mathcal{A}_{d_1}\mathcal{A}_{d_1}^{\dagger}
  +\mathcal{E}_{d_1},\\
  &\mathcal{H}_{d_1}\phi_{d_1\, n}(x)=\mathcal{E}_n\phi_{d_1\, n}(x),
  \quad\phi_{d_1\, n}(x)\eqdef\mathcal{A}_{d_1}\phi_n(x),\quad
  n\in\mathbb{Z}_+\backslash\{d_1\}.
\end{align}
Note that the energy level $d_1$ is now deleted,
$\phi_{d_1\, d_1}(x)\equiv0$, from the set of `eigenfunctions'
$\{\phi_{d_1\, n}(x)\}$ of the new Hamiltonian $\mathcal{H}_{d_1}$.

Suppose we have determined  $\mathcal{H}_{d_1\,\ldots\,d_s}$ and
$\phi_{d_1\,\ldots\,d_s\,n}(x)$ with $s$ deletions. They have the
following properties
\begin{align}
  &\mathcal{H}_{d_1\,\ldots\,d_s}
  \eqdef\mathcal{A}_{d_1\,\ldots\,d_s}
  \mathcal{A}_{d_1\,\ldots\,d_s}^{\dagger}+\mathcal{E}_{d_s},
  \label{dQMHb1bsdef}\\
  &\mathcal{A}_{d_1\,\ldots\,d_s}\eqdef
  i\Bigl(e^{\frac{\gamma}{2}p}\sqrt{V_{d_1\,\ldots\,d_s}^*(x)}
  -e^{-\frac{\gamma}{2}p}\sqrt{V_{d_1\,\ldots\,d_s}(x)}\,\Bigr),\n
  &\mathcal{A}_{d_1\,\ldots\,d_s}^{\dagger}\eqdef
  -i\Bigl(\sqrt{V_{d_1\,\ldots\,d_s}(x)}\,e^{\frac{\gamma}{2}p}
  -\sqrt{V_{d_1\,\ldots\,d_s}^*(x)}\,e^{-\frac{\gamma}{2}p}\Bigr),\\
  &V_{d_1\,\ldots\,d_s}(x)\eqdef
  \left\{\begin{array}{ll}
  {\displaystyle
  \sqrt{V_{d_1\,\ldots\,d_{s-1}}(x-i\tfrac{\gamma}{2})
  V_{d_1\,\ldots\,d_{s-1}}^*(x-i\tfrac{\gamma}{2})}\,
  \frac{\phi_{d_1\,\ldots\,d_s}(x-i\gamma)}
  {\phi_{d_1\,\ldots\,d_s}(x)}}&(s\geq 2),\\
  {\displaystyle
  \sqrt{V(x)V^*(x-i\gamma)}\,\frac{\phi_{d_1}(x-i\gamma)}{\phi_{d_1}(x)}}
  &(s=1),
  \end{array}\right.\\
  &\phi_{d_1\,\ldots\,d_s\,n}(x)
  \eqdef\mathcal{A}_{d_1\,\ldots\,d_s}\phi_{d_1\,\ldots\,d_{s-1}\,n}(x),
  \quad\phi_{d_1\,\ldots\,d_s\,n}(x)=\phi_{d_1\,\ldots\,d_s\,n}^*(x),
  \label{dQMphib1bsndef}\\
  &\mathcal{H}_{d_1\,\ldots\,d_s}\phi_{d_1\,\ldots\,d_s\,n}(x)
  =\mathcal{E}_n\phi_{d_1\,\ldots\,d_s\,n}(x),
  \label{dQMHb1bnphi=..}
\end{align}
where $n\in\mathbb{Z}_+\backslash\{d_1,\ldots,d_s\}$.
We have also
\begin{equation}
  \phi_{d_1\,\ldots\,d_{s-1}\,n}(x)
  =\frac{\mathcal{A}_{d_1\,\ldots\,d_s}^{\dagger}}
  {\mathcal{E}_n-\mathcal{E}_{d_s}}\phi_{d_1\,\ldots\,d_s\,n}(x)
  \quad(n\in \mathbb{Z}_+\backslash\{d_1,\ldots,d_s\}).
\end{equation}
Next we will define a new Hamiltonian system with one more deletion of
the level $d_{s+1}$.
We can show the following:
\begin{align}
  &\mathcal{H}_{d_1\,\ldots\,d_s}
  =\mathcal{A}_{d_1\,\ldots\,d_s\,d_{s+1}}^{\dagger}
  \mathcal{A}_{d_1\,\ldots\,d_s\,d_{s+1}}
  +\mathcal{E}_{d_{s+1}},\quad
  \mathcal{A}_{d_1\,\ldots\,d_s\,d_{s+1}}
  \phi_{d_1\,\ldots\,d_s\,d_{s+1}}(x)=0,
  \label{Hb1bs}\\
  &\mathcal{A}_{d_1\,\ldots\,d_s\,d_{s+1}}\eqdef
  i\Bigl(e^{\frac{\gamma}{2}p}\sqrt{V_{d_1\,\ldots\,d_s\,d_{s+1}}^*(x)}
  -e^{-\frac{\gamma}{2}p}\sqrt{V_{d_1\,\ldots\,d_s\,d_{s+1}}(x)}\,\Bigr),\n
  &\mathcal{A}_{d_1\,\ldots\,d_s\,d_{s+1}}^{\dagger}\eqdef
  -i\Bigl(\sqrt{V_{d_1\,\ldots\,d_s\,d_{s+1}}(x)}\,e^{\frac{\gamma}{2}p}
  -\sqrt{V_{d_1\,\ldots\,d_s\,d_{s+1}}^*(x)}\,e^{-\frac{\gamma}{2}p}\Bigr),\\
  &V_{d_1\,\ldots\,d_s,d_{s+1}}(x)\eqdef
  \sqrt{V_{d_1\,\ldots\,d_s}(x-i\tfrac{\gamma}{2})
  V_{d_1\,\ldots\,d_s}^*(x-i\tfrac{\gamma}{2})}\,
  \frac{\phi_{d_1\,\ldots\,d_s\,d_{s+1}}(x-i\gamma)}
  {\phi_{d_1\,\ldots\,d_s\,d_{s+1}}(x)}.
\end{align}
These determine a new Hamiltonian system with $s+1$ deletions:
\begin{align}
  &\mathcal{H}_{d_1\,\ldots\,d_{s+1}}
  \eqdef\mathcal{A}_{d_1\,\ldots\,d_{s+1}}
  \mathcal{A}_{d_1\,\ldots\,d_{s+1}}^{\dagger}+\mathcal{E}_{d_{s+1}},
  \label{d2QMHb1bsdef}\\
  &\phi_{d_1\,\ldots\,d_{s+1}\,n}(x)
  \eqdef\mathcal{A}_{d_1\,\ldots\,d_{s+1}}\phi_{d_1\,\ldots\,d_{s}\,n}(x),
  \quad\phi_{d_1\,\ldots\,d_{s+1}\,n}(x)=\phi_{d_1\,\ldots\,d_{s+1}\,n}^*(x),
  \label{d2QMphib1bsndef}\\
  &\mathcal{H}_{d_1\,\ldots\,d_{s+1}}\phi_{d_1\,\ldots\,d_{s+1}\,n}(x)
  =\mathcal{E}_n\phi_{d_1\,\ldots\,d_{s+1}\,n}(x),
  \label{d2QMHb1bnphi=..}
\end{align}
where $n\in\mathbb{Z}_+\backslash\{d_1,\ldots,d_{s+1}\}$.

\bigskip
After deleting all the $\mathcal{D}=\{d_1,\,\cdots,\,d_\ell\}$ energy
levels, the resulting Hamiltonian system
$\mathcal{H}_{\mathcal{D}}\equiv\mathcal{H}_{d_1\,\ldots\,d_{\ell}}$,
$\mathcal{A}_{\mathcal{D}}\equiv\mathcal{A}_{d_1\,\ldots\,d_{\ell}}$,
etc has the following form:
\begin{align}
  &\mathcal{H}_{\mathcal{D}}
  \eqdef\mathcal{A}_{\mathcal{D}}
  \mathcal{A}_{\mathcal{D}}^{\dagger}+\mathcal{E}_{d_{\ell}},
  \label{2dfQMHb1bsdef}\\
  &\mathcal{A}_{\mathcal{D}}\eqdef
  i\Bigl(e^{\frac{\gamma}{2}p}\sqrt{V_{\mathcal{D}}^*(x)}
  -e^{-\frac{\gamma}{2}p}\sqrt{V_{\mathcal{D}}(x)}\,\Bigr),\quad
  \mathcal{A}_{\mathcal{D}}^{\dagger}\eqdef
  -i\Bigl(\sqrt{V_{\mathcal{D}}(x)}\,e^{\frac{\gamma}{2}p}
  -\sqrt{V_{\mathcal{D}}^*(x)}\,e^{-\frac{\gamma}{2}p}\Bigr),\\
  &V_{\mathcal{D}}(x)\eqdef
  \sqrt{V_{d_1\,\ldots\,d_{\ell-1}}(x-i\tfrac{\gamma}{2})
  V_{d_1\,\ldots\,d_{\ell-1}}^*(x-i\tfrac{\gamma}{2})}\,
  \frac{\phi_{\mathcal{D}}(x-i\gamma)}
  {\phi_{\mathcal{D}}(x)},\\
  &\phi_{{\mathcal{D}}\,n}(x)
  \eqdef\mathcal{A}_{\mathcal{D}}\phi_{d_1\,\ldots\,d_{\ell-1}\,n}(x),
  \quad\phi_{{\mathcal{D}}\,n}(x)=\phi_{{\mathcal{D}}\,n}^*(x),
  \quad(n\in\mathbb{Z}_+\backslash{\mathcal{D}}),
  \label{2dQMphib1bsndef}\\
  &\mathcal{H}_{\mathcal{D}}\phi_{{\mathcal{D}}\,n}(x)
  =\mathcal{E}_n\phi_{{\mathcal{D}}\,n}(x)
  \quad(n\in\mathbb{Z}_+\backslash{\mathcal{D}}).
  \label{2dfQMHb1bnphi=..}
\end{align}
Now that $\mathcal{H}_{\mathcal{D}}$ has the lowest energy level $\mu$:
\begin{equation}
  \mu\eqdef \min\{n\,|\,n\in \mathbb{Z}_+\backslash{\mathcal{D}}\},
\end{equation}
with the groundstate wavefunction $\bar{\phi}_\mu(x)$
\begin{equation}
  \bar{\phi}_\mu(x)\eqdef\phi_{{\mathcal{D}}\,\mu}(x)\equiv
  \phi_{d_1\,\cdots\,d_\ell\, \mu}(x).
\end{equation}
Then the Hamiltonian system can be expressed simply in terms of the
groundstate wavefunction $\bar{\phi}_\mu(x)$, which we will denote
by new symbols $\bar{\mathcal{H}}$, $\bar{\mathcal{A}}$, etc:
\begin{align}
  &\bar{\mathcal{H}}\equiv\mathcal{H}_{\mathcal{D}}
  \eqdef\bar{\mathcal{A}}^\dagger
  \bar{\mathcal{A}}+\mathcal{E}_{\mu},
  \qquad
  \bar{\mathcal{A}}\bar{\phi}_{\mu}(x)=0,
  \label{dbfQMHb1bsdef}\\
  &\bar{\mathcal{A}}\equiv \mathcal{A}_{\mathcal{D}\,\mu}\eqdef
  i\Bigl(e^{\frac{\gamma}{2}p}\sqrt{\bar{V}^*(x)}
  -e^{-\frac{\gamma}{2}p}\sqrt{\bar{V}(x)}\,\Bigr),\n
  &\bar{\mathcal{A}}^{\dagger}\equiv
  \mathcal{A}_{\mathcal{D}\,\mu}^\dagger\eqdef
  -i\Bigl(\sqrt{\bar{V}(x)}\,e^{\frac{\gamma}{2}p}
  -\sqrt{\bar{V}^*(x)}\,e^{-\frac{\gamma}{2}p}\Bigr),\\
  &\bar{V}(x)\equiv V_{\mathcal{D}\,\mu}(x)\eqdef
  \sqrt{V_{\mathcal{D}}(x-i\tfrac{\gamma}{2})
  V_{\mathcal{D}}^*(x-i\tfrac{\gamma}{2})}\,
  \frac{\bar{\phi}_{\mu}(x-i\gamma)}
  {\bar{\phi}_{\mu}(x)},\\
  &\bar{\mathcal{H}}\bar{\phi}_{n}(x)
  =\mathcal{E}_n\bar{\phi}_{n}(x),\quad
  \bar{\phi}_{n}(x)\equiv\phi_{{\mathcal{D}}\,n}(x)
  \qquad(n\in\mathbb{Z}_+\backslash{\mathcal{D}}).
  \label{dbfQMHb1bnphi=..}
\end{align}

The discrete counterpart of the determinant formulas
\eqref{QMphib1bsn}--\eqref{Ub1bs} requires a deformation of the Wronskian,
the Casorati determinant, which has a good limiting property:
\begin{align}
  &\text{W}_{\gamma}[f_1,\ldots,f_n](x)
  \eqdef i^{\frac12n(n-1)}
  \det\Bigl(f_k(x+i\tfrac{n+1-2j}{2}\gamma)\Bigr)_{1\leq j,k\leq n},\\
  &\lim_{\gamma\to 0}\gamma^{-\frac12n(n-1)}
  \text{W}_{\gamma}[f_1,f_2,\ldots,f_n](x)
  =\text{W}\,[f_1,f_2,\ldots,f_n](x),
\end{align}
(for $n=0$, we set $\text{W}_{\gamma}\,[\cdot](x)=1$.).
It satisfies
\begin{align}
  &\text{W}_{\gamma}[f_1,\ldots,f_n]^*(x)
  =\text{W}_{\gamma}[f_1^*,\ldots,f_n^*](x),\\
  &\text{W}_{\gamma}[gf_1,gf_2,\ldots,gf_n]
  =\prod_{j=1}^ng(x+i\tfrac{n+1-2j}{2}\gamma)\cdot
  \text{W}_{\gamma}[f_1,f_2,\ldots,f_n](x),
  \label{dWformula1}\\
  &\text{W}_{\gamma}\bigl[\text{W}_{\gamma}[f_1,f_2,\ldots,f_n,g],
  \text{W}_{\gamma}[f_1,f_2,\ldots,f_n,h]\,\bigr](x)\n
  &=\text{W}_{\gamma}[f_1,f_2,\ldots,f_n](x)\,
  \text{W}_{\gamma}[f_1,f_2,\ldots,f_n,g,h](x)
  \quad(n\geq 0).
  \label{dWformula2}
\end{align}
By using the Casorati determinant we obtain ($\ell\geq 0$)
\begin{align}
  &\bar{\phi}_n(x)\equiv\phi_{d_1\,\ldots\,d_\ell\,n}(x)=
  \sqrt{\prod_{j=1}^\ell
  V_{d_1\,\ldots\,d_j}(x+i\tfrac{\ell+1-j}{2}\gamma)}\,\,
  \frac{\text{W}_{\gamma}\,[\phi_{d_1},\ldots,\phi_{d_\ell},\phi_n](x)}
  {\text{W}_{\gamma}\,[\phi_{d_1},\ldots,\phi_{d_\ell}]
  (x-i\tfrac{\gamma}{2})},
  \label{phib1bsn}\\
  &\bar{V}(x)\equiv V_{d_1\,\ldots\,d_\ell\,\mu}(x)=
  \sqrt{V(x-i\tfrac{\ell}{2}\gamma)V^*(x-i\tfrac{\ell+2}{2}\gamma)}\n
  &\phantom{V_{d_1\,\ldots\,d_\ell\,\mu}(x)=}\times
  \frac{\text{W}_{\gamma}\,[\phi_{d_1},\ldots,\phi_{d_\ell}]
  (x+i\tfrac{\gamma}{2})}
  {\text{W}_{\gamma}\,[\phi_{d_1},\ldots,\phi_{d_\ell}]
  (x-i\tfrac{\gamma}{2})}\,
  \frac{\text{W}_{\gamma}\,[\phi_{d_1},\ldots,\phi_{d_{\ell}},\phi_{\mu}]
  (x-i\gamma)}
  {\text{W}_{\gamma}\,[\phi_{d_1},\ldots,\phi_{d_{\ell}},\phi_{\mu}](x)}.
  \label{Vb1bs+1}
\end{align}
We also have ($\ell\geq 0$)
\begin{equation}
  \prod_{j=1}^\ell V_{d_1\,\ldots\,d_j}(x+i\tfrac{\ell+1-j}{2}\gamma)
  =\sqrt{\prod_{j=0}^{\ell-1}V(x+i\tfrac{\ell-2j}{2}\gamma)
  V^*(x-i\tfrac{\ell-2j}{2}\gamma)}\,\,
  \frac{\text{W}_{\gamma}\,[\phi_{d_1},\ldots,\phi_{d_\ell}]
  (x-i\tfrac{\gamma}{2})}
  {\text{W}_{\gamma}\,[\phi_{d_1},\ldots,\phi_{d_\ell}]
  (x+i\tfrac{\gamma}{2})}.
  \label{prodV}
\end{equation}
Therefore $V_{d_1\,\ldots\,d_\ell}(x)$ and
$\phi_{d_1,\ldots\,d_\ell\,n}(x)$ are symmetric with respect to
$d_1,\ldots,d_\ell$, and $\mathcal{H}_{d_1\,\ldots\,d_\ell}$ is independent
of the order of $\{d_j\}$.

Let us state the discrete QM analogue of Adler's theorem;
{\em If the set of deleted energy levels $\mathcal{D}=\{d_1,\ldots,d_\ell\}$
satisfy the condition \eqref{dellcond}, the modified Hamiltonian is given
by $\bar{\mathcal{H}}=\mathcal{H}_{d_1\,\ldots\,d_\ell}
=\bar{\mathcal{A}}^{\dagger}\bar{\mathcal{A}}+\mathcal{E}_{\mu}$
with the potential function given by \eqref{Vb1bs+1} and its eigenfunctions
are given by\/} \eqref{phib1bsn}.
The discrete QM version of Crum's theorem  \cite{os15} corresponds to
the choice $\{d_1,\ldots,d_\ell\}=\{0,1,\ldots,\ell-1\}$ and the new
groundstate is at the level $\mu=\ell$ and there is no vacant energy
level above that.
Due to the lack of generic theorems in the theory of difference equations,
the hermiticity of the resulting Hamiltonian $\bar{\mathcal{H}}$ and the
non-singularity of the eigenfunctions $\bar{\phi}_n(x)$ cannot be proved
categorically for the discrete QM, even when the condition \eqref{dellcond}
is satisfied by the deleted levels.
See Appendix A of \cite{os13} for a detailed discussion of the
self-adjointness of the Hamiltonians in discrete QM.
It should be stressed that in most practical cases, in particular,
in the cases of polynomial eigenfunctions, the hermiticity of the
Hamiltonian $\bar{\mathcal{H}}$ and non-singularity of the eigenfunctions
$\{\bar{\phi}_n(x)\}$ are satisfied.

\subsection{Polynomial eigenfunctions}

In this subsection we consider the typical case of shape invariant
systems in which the eigenfunctions consist of the orthogonal polynomials
$\{P_n\}$:
\begin{equation}
  \phi_n(x)=\phi_0(x)P_n(\eta(x)),\quad\mathcal{A}\phi_0(x)=0,
\end{equation}
in which $\eta(x)$ is called the {\em sinusoidal coordinate\/}, see
\eqref{dsinu}. The groundstate wavefunction $\phi_0(x)$ provides the
orthogonality weight function
\begin{align}
  &\int_{x_1}^{x_2}\!\!\phi_{0}(x)^2\,P_n(\eta(x))P_m(\eta(x))dx
  =h_n\delta_{nm},\quad n,\,m\in\mathbb{Z}_+.
\end{align}
The difference equation for $\{P_n\}$ looks much simpler than the
Schr\"odinger equation \eqref{difschro}:
\begin{align}
  V(x)\bigl(P_n(\eta(x-i\gamma))-P_n(\eta(x))\bigr)
  +V^*(x)\bigl(P_n(\eta(x+i\gamma))-P_n(\eta(x))\bigr)
  =\mathcal{E}_nP_n(\eta(x)).
  \label{polyeq}
\end{align}
For the explicit forms of $V(x)$, see for example \eqref{Vform},
these are the equations that determine the hypergeometric orthogonal
polynomials, {\em e.g.\/} the Meixner-Pollaczek (MP), the continuous
Hahn (cH), the Wilson (W) and the Askey-Wilson (AW) polynomials.
In fact, the above form of the {\em difference equation} \eqref{polyeq}
is independent of the fact that $P_n$ is a polynomial or not.
It is obtained simply by the similarity transformation of the Hamiltonian
\eqref{discrham} in terms of the groundstate wavefunction $\phi_0(x)$:
\begin{align}
  \widetilde{\mathcal{H}}&\eqdef\phi_0(x)^{-1}\circ\mathcal{H}\circ
  \phi_0(x)=V(x)\,e^{\gamma p}+V(x)^*\,e^{-\gamma p}-V(x)-V(x)^*.
\end{align}

In the case of polynomial eigenfunctions, the modification of Crum's
theorem produces the eigenfunctions $\{\bar{\phi}_n(x)\}$ which again
consist of polynomials in $\eta(x)$. By using the property
\eqref{dWformula1} we have
\begin{align}
  &\text{W}_{\gamma}[\phi_1,\ldots,\phi_{\ell}](x)
  =\prod_{j=1}^{\ell}
  \phi_0\bigl(x+i\tfrac{\ell+1-2j}{2}\gamma\bigr)\cdot
  \text{W}_{\gamma}[\check{P}_1,\ldots,\check{P}_{\ell}](x),\\
  &\text{W}_{\gamma}[\phi_1,\ldots,\phi_{\ell},\phi_n](x)
  =\prod_{j=1}^{\ell+1}
  \phi_0\bigl(x+i\tfrac{\ell+2-2j}{2}\gamma\bigr)\cdot
  \text{W}_{\gamma}[\check{P}_1,\ldots,\check{P}_{\ell},\check{P}_n](x),
\end{align}
where $\check{P}_n(x)\eqdef P_n(\eta(x))$. Corresponding to the formula
\eqref{Wformula3} in the ordinary QM, we have 
\begin{align}
  &\check{f}_j(x)\eqdef f_j(\eta(x)),\quad
  f_j(\eta)\text{: polynomial in $\eta$},\n
  &\text{W}_{\gamma}[\check{f}_1,\check{f}_2,\ldots,\check{f}_n](x)
  =\varphi_n(x)\times\bigl(\text{polynomial in $\eta(x)$}\bigr),
\end{align}
in which $\varphi_n(x)$ is defined in \eqref{varphildef}
(and $\varphi(x)$ is defined in \eqref{varphidef}).

These simply mean that the resulting eigenfunctions $\{\bar{\phi}_n(x)\}$
\eqref{phib1bsn} are again polynomials in $\eta(x)$:
\begin{align}
  \bar{\phi}_n(x)&=\bar{\psi}(x)\,\mathcal{P}_n(\eta(x)),\\
  \bar{\psi}(x)&\eqdef
  \sqrt{\prod_{j=1}^\ell V_{d_1\,\ldots\,d_j}(x+i\tfrac{\ell+1-j}{2}\gamma)}
  \,\frac{\varphi_{\ell+1}(x)}{\varphi_{\ell}(x-i\frac{\gamma}{2})}
  \frac{\phi_0(x+i\tfrac{\ell}{2}\gamma)}
  {\mathcal{Q}(\eta(x-i\tfrac{\gamma}{2}))}\n
  &=\sqrt{\prod_{k=0}^{\ell-1}\varphi(x-i\tfrac{k}{2}\gamma)
  \sqrt{V(x+i\tfrac{\ell-2k}{2}\gamma)}\cdot
  \frac{\phi_0(x-i\frac{\ell}{2}\gamma)}
  {\mathcal{Q}(\eta(x-i\tfrac{\gamma}{2}))}}\n
  &\phantom{=}\times
  \sqrt{\prod_{k=0}^{\ell-1}\varphi(x+i\tfrac{k}{2}\gamma)
  \sqrt{V^*(x-i\tfrac{\ell-2k}{2}\gamma)}\cdot
  \frac{\phi_0(x+i\frac{\ell}{2}\gamma)}
  {\mathcal{Q}(\eta(x+i\tfrac{\gamma}{2}))}}\,,
\end{align}
in which $\mathcal{P}_n(\eta(x))$ and $\mathcal{Q}(\eta(x))$ are certain
polynomials in $\eta(x)$ defined by
\begin{equation}
  \text{W}_{\gamma}[\check{P}_{d_1},\ldots,\check{P}_{d_\ell},\check{P}_n]
  =\varphi_{\ell+1}(x)\times\mathcal{P}_n(\eta(x)),\quad
  \text{W}_{\gamma}[\check{P}_{d_1},\ldots,\check{P}_{d_\ell}]
  =\varphi_{\ell}(x)\times\mathcal{Q}(\eta(x)).
  \label{calPQdef}
\end{equation}
The polynomials $\{\mathcal{P}_n\}$ form a complete basis of the Hilbert
space and satisfy the orthogonality relations
\begin{equation}
  \int_{x_1}^{x_2}\bar{\psi}(x)^2\,
  \mathcal{P}_n(\eta(x))\mathcal{P}_m(\eta(x))dx
  =\bar{h}_n\delta_{nm},\quad n,\,m\in\mathbb{Z}_+\backslash\mathcal{D}.
\end{equation}
Let us emphasise that $n$ is the level of the original eigenfunction and
not the {\em degree} in $\eta$. The degree of $\mathcal{P}_n(\eta)$ depends
on the set $\mathcal{D}$, and it can be calculated explicitly from
\eqref{calPQdef}.
By construction $\ell$ members are missing:
$\mathcal{P}_{d_1}=\mathcal{P}_{d_2}=\cdots=\mathcal{P}_{d_\ell}\equiv0$.
Therefore these polynomials cannot be called {\em exceptional orthogonal
polynomials} \cite{gomez,os16,os19}.

\section{Summary and Comments}
\label{summary}
\setcounter{equation}{0}

Theory of exactly solvable discrete QM is less developed than that of
the ordinary QM. Up to date, the known exactly solvable discrete quantum
systems are all shape invariant \cite{os12,os13} and in one to one
correspondence with the known ($q$)-hypergeometric orthogonal polynomials
\cite{nikiforov}--\cite{koeswart}.
Small progress was made in this direction \cite{os14} by introducing
several new sinusoidal coordinates for the construction of new types
of exactly solvable Hamiltonians.
Roughly speaking, this approach attempts to create the discrete analogues
of various Morse type potentials and the soliton potentials.
In this paper we pursue another direction; to construct infinitely many
exactly solvable quantum systems by {\em deforming\/} the known exactly
solvable one.
In the ordinary QM, the modification of Crum's theorem \cite{crum} due to
Krein-Adler \cite{adler} allows to produce an essentially iso-spectral
Hamiltonian
by deleting a finite number of energy levels from the original system.
The set of deleted level must satisfy certain condition \eqref{dellcond},
but there are infinitely many possible deletions leading to infinitely
many exactly solvable systems starting from a known one. The discrete
analogue of Adler's modification is presented in this paper in parallel
with the original version, since the algebraic structure is common.
We also comment on the practical cases when the eigenfunctions consist
of orthogonal polynomials. The eigenfunctions of the resulting system
also consist of orthogonal polynomials. But certain members of these
polynomials are missing due to the deletion.

Very special and simple examples, in which all the excited states from
the first to the $\ell$-th are deleted (see Fig.\,2), are presented
explicitly in Appendix.
As will be commented shortly, these examples were instrumental for the
discovery of the infinitely many shape invariant systems and the
corresponding infinitely many exceptional orthogonal polynomials
\cite{os16,os17}.
In the ordinary QM, the corresponding prepotential has a very simple
form \eqref{defellprep}:
\begin{equation*}
  w_{\ell}(x;\bm{\lambda})=
  \mathcal{W}(x;\bm{\lambda}+\ell\bm{\delta})
  +\log\frac{1}{\xi_{\ell}(\eta(x);\bm{\lambda})},
\end{equation*}
to be compared with the prepotential for the Hamiltonian of the $\ell$-th
exceptional orthogonal polynomial (14) and (28) of \cite{os16}:
\begin{equation}
  w_{\ell}(x;\bm{\lambda})=
  \mathcal{W}(x;\bm{\lambda}+\ell\bm{\delta})
  +\log\frac{\xi_{\ell}(\eta(x);\bm{\lambda}+\bm{\delta})}
  {\xi_{\ell}(\eta(x);\bm{\lambda})}.
\end{equation}
In the discrete QM, the corresponding formula is \eqref{vellexp2}
\begin{equation*}
  V_{\ell}(x;\bm{\lambda})=\kappa^\ell
  \frac{\xi_{\ell}(\eta(x+i\frac{\gamma}{2});\bm{\lambda})}
  {\xi_{\ell}(\eta(x-i\frac{\gamma}{2});\bm{\lambda})}
  V(x;\bm{\lambda}+\ell\bm{\delta}),
\end{equation*}
to be compared with the corresponding formula for the Hamiltonian of
the $\ell$-th exceptional orthogonal polynomial (30)  of \cite{os17}:
\begin{equation}
  V_{\ell}(x;\bm{\lambda})=
  \frac{\xi_{\ell}(\eta(x-i\gamma);\bm{\lambda}+\bm{\delta})}
  {\xi_{\ell}(\eta(x);\bm{\lambda}+\bm{\delta})}
  \frac{\xi_{\ell}(\eta(x+i\frac{\gamma}{2});\bm{\lambda})}
  {\xi_{\ell}(\eta(x-i\frac{\gamma}{2});\bm{\lambda})}
  V(x;\bm{\lambda}+\ell\bm{\delta}).
\end{equation}
The addition (multiplication) of the deforming polynomial with the
shifted parameters $\xi_{\ell}(\eta(x);\bm{\lambda}+\bm{\delta})$ would
achieve the shape invariance.
Since the harmonic oscillator has no shiftable parameter, we have
$\xi_{\ell}(\eta(x);\bm{\lambda})
=\xi_{\ell}(\eta(x);\bm{\lambda}+\bm{\delta})$. This also `explains'
non-existence of  exceptional Hermite polynomials. In contrast to the
Hermite polynomial, the continuous Hahn polynomial has four real parameters.
We can construct the corresponding exceptional continuous Hahn polynomials
with three real parameters, which will be reported elsewhere. 

The actual function
forms of the deforming polynomial $\xi_{\ell}$ in Appendix are not the
same as those for the exceptional orthogonal polynomials.
For the ordinary QM examples, see \eqref{QMxil} vs. (13) and (27) in
\cite{os16} and for the discrete QM examples, see \eqref{dQMxil} vs.
(64) and (78) in \cite{os17}. But they share some interesting features.

Before closing this section, let us remark that the present modification
of Crum's theorem is applicable to the Hamiltonian systems of various
species of the infinite family of exceptional orthogonal polynomials
\cite{os16, os17, os19}, as well as to those of the classical orthogonal
polynomials including the Wilson and the Askey-Wilson polynomials.

\section*{Acknowledgements}

We thank Reido Kobayashi for useful comments and discussion.
R.\,S. is supported in part by Grant-in-Aid for Scientific Research from
the Ministry of Education, Culture, Sports, Science and Technology (MEXT),
No.19540179.
L.\,G-G. thanks MEXT for the research studentship.

\bigskip
\appendix
\renewcommand{\theequation}{\Alph{section}.\arabic{equation}}
\setcounter{equation}{0}

\noindent
{\Large\bf Appendix}

\noindent
In Appendix we present very special and simple examples of an application
of Adler's theorem, in which the eigenstates
$\phi_1,\phi_2,\ldots,\phi_{\ell}$ are deleted. In other words,
$\mathcal{D}=\{d_1,d_2,\ldots,d_\ell\}$$=\{1,2,\ldots,\ell\}$, that is,
the modified groundstate level is the same as that of the original
theory $\mu=0$.
The situation is illustrated in Fig.\,2, which should be compared with
Fig.\,1, depicting the generic case discussed in sections \ref{genordQM}
and \ref{gendQM}.

\bigskip
\begin{center}
  \includegraphics{adlerscheme.epsi}
  \hspace*{20mm}
  \includegraphics{adlerscheme2.epsi}
\end{center}
\vspace*{-3mm}
\hspace*{14mm}
Figure\,1: Generic case
\hspace*{40mm}
Figure\,2: Special case

\bigskip
The black circles denote the energy levels, whereas the white circles
denote {\em deleted\/} energy levels.
We write $\bar{\mathcal{H}}=\mathcal{H}_{12\,\ldots\,\ell}$,
$\bar{\phi}_n=\phi_{12\,\ldots\,\ell\,n}$,
$\bar{\mathcal{A}}=\mathcal{A}_{12\,\ldots\,\ell}$,
$\bar{V}=V_{12\,\ldots\,\ell}$ etc.\ as $\mathcal{H}_{\ell}$,
$\phi_{\ell,n}$, $\mathcal{A}_{\ell}$, $V_{\ell}$ etc.
This Hamiltonian $\mathcal{H}_{\ell}
=\mathcal{A}_{\ell}^\dagger\mathcal{A}_{\ell}$ is non-singular for even
$\ell$ but may be singular for odd $\ell$.
Since algebraic formulas such as the Wronskians and Casoratians are
valid for even and odd $\ell$, we present various formulas without
restricting to the even $\ell$. The original systems are shape invariant
but the $(\phi_1,\ldots,\phi_{\ell})$-deleted systems $\mathcal{H}_{\ell}$
are not.
The rightmost vertical line in Fig.\,2 corresponds to the Hamiltonian
system $\mathcal{H}_{\ell}'=\mathcal{A}_{\ell}\mathcal{A}_{\ell}^\dagger$,
which is shape invariant and it is obtained from $\mathcal{H}_{\ell}$ by
one more step of Crum's method. This study helped us to find the new
shape invariant systems and exceptional orthogonal polynomials
\cite{os16,os17}.

\section{The ordinary QM}

Here we apply Adler's theorem to the harmonic oscillator, the radial
oscillator and the Darboux-P\"{o}schl-Teller potential, whose
eigenfunctions are described by the classical orthogonal polynomials.
That is, the Hermite, Laguerre and Jacobi polynomials, to be abbreviated
as H, L and J, respectively.
These original systems are shape invariant, meaning a very special
form of parameter dependence, \eqref{shapeinv}, \eqref{shapeinv2}.
Here we display the parameter dependence explicitly by $\bm{\lambda}$,
which represents the set of the parameters.

\subsection{The original systems}

Here we summarise various properties of the original Hamiltonian systems
to be compared with the specially modified systems to be presented in
\ref{sec:A2}.
Let us start with the Hamiltonians, Schr\"{o}dinger equations and
eigenfunctions ($x_1<x<x_2$):
\begin{align}
  &\mathcal{H}(\bm{\lambda})\eqdef
  \mathcal{A}(\bm{\lambda})^{\dagger}\mathcal{A}(\bm{\lambda}),\quad
  \mathcal{A}(\bm{\lambda})\eqdef
  \frac{d}{dx}-\frac{d\mathcal{W}(x;\bm{\lambda})}{dx},\quad
  \mathcal{A}(\bm{\lambda})^{\dagger}
  =-\frac{d}{dx}-\frac{d\mathcal{W}(x;\bm{\lambda})}{dx},\\
  &\mathcal{H}(\bm{\lambda})\phi_n(x;\bm{\lambda})
  =\mathcal{E}_n(\bm{\lambda})\phi_n(x;\bm{\lambda})\quad
  (n=0,1,2,\ldots),\\
  &\phi_n(x;\bm{\lambda})=\phi_0(x;\bm{\lambda})P_n(\eta(x);\bm{\lambda}),
  \quad \phi_0(x;\bm{\lambda})=e^{\mathcal{W}(x;\bm{\lambda})}.
\end{align}
Here $\eta(x)$ is the sinusoidal coordinate, $\mathcal{W}(x;\bm{\lambda})$
is the prepotential and $\mathcal{E}_n(\bm{\lambda})$ is the $n$-th energy
eigenvalue:
\begin{align}
  &\eta(x)\eqdef\left\{
  \begin{array}{llll}
  x,& x_1=-\infty,& x_2=\infty,&:\text{H}\\
  x^2,& x_1=0, &x_2=\infty,&:\text{L}\\
  \cos 2x,&x_1=0,&x_2=\tfrac{\pi}{2},&:\text{J}
  \end{array}\right.\!\!,
  \quad
  \bm{\lambda}\eqdef\left\{
  \begin{array}{lll}
  \text{none}&&:\text{H}\\
  g,&g>0&:\text{L}\\
  (g,h),&g,h>0&:\text{J}
  \end{array}\right.\!\!,\\
  &\mathcal{W}(x;\bm{\lambda})\eqdef\left\{
  \begin{array}{ll}
  -\frac12x^2&:\text{H}\\[1pt]
  -\frac12x^2+g\log x&:\text{L}\\[1pt]
  g\log\sin x+h\log\cos x&:\text{J}
  \end{array}\right.\!\!,
  \quad
  \mathcal{E}_n(\bm{\lambda})\eqdef\left\{
  \begin{array}{ll}
  2n&:\text{H}\\
  4n&:\text{L}\\
  4n(n+g+h)&:\text{J}
  \end{array}\right.\!\!.
\end{align}
The eigenfunction consists of an orthogonal polynomial
$P_n(\eta;\bm{\lambda})$, a polynomial of degree $n$ in $\eta$,
($P_n(\eta;\bm{\lambda})=0$ for $n<0$):
\begin{align}
  &P_n(\eta;\bm{\lambda})\eqdef c_n(\bm{\lambda})
  P_n^{\text{monic}}(\eta;\bm{\lambda}),\\
  &P_n(\eta;\bm{\lambda})\eqdef\left\{
  \begin{array}{ll}
  H_n(\eta)&:\text{H}\\[1pt]
  L_n^{(g-\frac12)}(\eta)&:\text{L}\\[1pt]
  P_n^{(g-\frac12,h-\frac12)}(\eta)&:\text{J}
  \end{array}\right.\!\!,
  \quad
  c_n(\bm{\lambda})\eqdef\left\{
  \begin{array}{ll}
  2^n&:\text{H}\\
  \frac{(-1)^n}{n!}&:\text{L}\\[1pt]
  \frac{(n+g+h)_n}{2^nn!}&:\text{J}
  \end{array}\right.\!\!,
\end{align}
in which $(a)_n$ is the Pochhammer symbol.
Shape invariance means
\begin{equation}
  \mathcal{A}(\bm{\lambda})\mathcal{A}(\bm{\lambda})^{\dagger}
  =\mathcal{A}(\bm{\lambda+\bm{\delta}})^{\dagger}
  \mathcal{A}(\bm{\lambda}+\bm{\delta})
  +\mathcal{E}_1(\bm{\lambda}),\quad
 \bm{\delta}\eqdef\left\{
  \begin{array}{ll}
  \text{none}&:\text{H}\\
  1&:\text{L}\\
  (1,1)&:\text{J}
  \end{array}\right.\!\!,
  \label{shapeinv}
\end{equation}
or equivalently,
\begin{equation}
  \Bigl(\frac{d\mathcal{W}(x;\bm{\lambda})}{dx}\Bigr)^2
  -\frac{d^2\mathcal{W}(x;\bm{\lambda})}{dx^2}
  =\Bigl(\frac{d\mathcal{W}(x;\bm{\lambda}+\bm{\delta})}{dx}\Bigr)^2
  +\frac{d^2\mathcal{W}(x;\bm{\lambda}+\bm{\delta})}{dx^2}
  +\mathcal{E}_1(\bm{\lambda}).
    \label{shapeinv2}
\end{equation}
The action of $\mathcal{A}(\bm{\lambda})$ and
$\mathcal{A}(\bm{\lambda})^{\dagger}$ on the eigenfunction is:
\begin{equation}
  \mathcal{A}(\bm{\lambda})\phi_n(x;\bm{\lambda})
  =f_n(\bm{\lambda})
  \phi_{n-1}\bigl(x;\bm{\lambda}+\bm{\delta}\bigr),\quad
  \mathcal{A}(\bm{\lambda})^{\dagger}
  \phi_{n-1}\bigl(x;\bm{\lambda}+\bm{\delta}\bigr)
  =b_{n-1}(\bm{\lambda})\phi_n(x;\bm{\lambda}).
\end{equation}
Here the coefficients $f_n(\bm{\lambda})$ and $b_{n-1}(\bm{\lambda})$ are
the factors of $\mathcal{E}_n(\bm{\lambda})$:
\begin{equation}
  f_n(\bm{\lambda})\eqdef\left\{
  \begin{array}{ll}
  2n&:\text{H}\\
  -2&:\text{L}\\
  -2(n+g+h)&:\text{J}
  \end{array}\right.\!\!,
  \quad
  b_{n-1}(\bm{\lambda})\eqdef\left\{
  \begin{array}{ll}
  1&:\text{H}\\
  -2n&:\text{L,\,J}
  \end{array}\right.\!\!,
  \quad \mathcal{E}_n(\bm{\lambda})=f_n(\bm{\lambda})b_{n-1}(\bm{\lambda}).
\end{equation}
The forward and backward shift operators, $\mathcal{F}(\bm{\lambda})$ and
$\mathcal{B}(\bm{\lambda})$, are defined by:
\begin{align}
  &\mathcal{F}(\bm{\lambda})\eqdef
  \phi_0(x;\bm{\lambda}+\bm{\delta})^{-1}\circ
  \mathcal{A}(\bm{\lambda})\circ\phi_0(x;\bm{\lambda})
  =\frac{\phi_0(x;\bm{\lambda})}
  {\phi_0(x;\bm{\lambda}+\bm{\delta})}\,\frac{d}{dx}\,,\\
  &\mathcal{B}(\bm{\lambda})\eqdef
  \phi_0(x;\bm{\lambda})^{-1}\circ
  \mathcal{A}(\bm{\lambda})^{\dagger}
  \circ\phi_0(x;\bm{\lambda}+\bm{\delta})\n
  &\phantom{\mathcal{B}(\bm{\lambda})}
  =-\frac{\phi_0(x;\bm{\lambda}+\bm{\delta})}{\phi_0(x;\bm{\lambda})}
  \Bigl(\frac{d}{dx}+\partial_x\bigl(\mathcal{W}(x;\bm{\lambda})
  +\mathcal{W}(x;\bm{\lambda}+\bm{\delta})\bigr)\Bigr),
\end{align}
and their action on the polynomial is:
\begin{align}
  &\mathcal{F}(\bm{\lambda})P_n(\eta(x);\bm{\lambda})
  =f_n(\bm{\lambda})P_{n-1}(\eta(x);\bm{\lambda}+\bm{\delta}),\\
  &\mathcal{B}(\bm{\lambda})P_{n-1}(\eta(x);\bm{\lambda}+\bm{\delta})
  =b_{n-1}(\bm{\lambda})P_n(\eta(x);\bm{\lambda}).
\end{align}
Note that $\mathcal{F}(\bm{\lambda})$ and $\mathcal{B}(\bm{\lambda})$
can also be expressed in terms of $\eta$ only \cite{hos}.
The orthogonality reads
\begin{align}
  &\int_{x_1}^{x_2}\!\!\phi_{0}(x;\bm{\lambda})^2\,
  P_n(\eta(x);\bm{\lambda})P_m(\eta(x);\bm{\lambda})dx
  =h_n(\bm{\lambda})\delta_{nm},\\
  &\qquad\quad h_n(\bm{\lambda})\eqdef\left\{
  \begin{array}{ll}
  2^nn!\sqrt{\pi}&:\text{H}\\[2pt]
  \frac{1}{2\,n!}\Gamma(n+g+\frac12)&:\text{L}\\[4pt]
  \frac{\Gamma(n+g+\frac12)\Gamma(n+h+\frac12)}
  {2\,n!(2n+g+h)\Gamma(n+g+h)}&:\text{J}
  \end{array}\right.\!\!.
\end{align}

\subsection{The $(\phi_1,\ldots,\phi_{\ell})$-deleted systems}
\label{sec:A2}

The prepotential of the modified system is obtained from \eqref{polyphi0}
up to an additive constant:
\begin{equation}
  w_{\ell}(x;\bm{\lambda})\eqdef\log\phi_{\ell,0}(x)=
  \mathcal{W}(x;\bm{\lambda}+\ell\bm{\delta})
  -\log\xi_{\ell}(\eta(x);\bm{\lambda}).
  \label{defellprep}
\end{equation}
It is a polynomial ($\xi_{\ell}(\eta(x);\bm{\lambda})$) deformation of
the shape invariant one $\mathcal{W}(x;\bm{\lambda}+\ell\bm{\delta})$.
Note that the normalization of $\xi_{\ell}$ does not affect the Hamiltonian.
The explicit forms of the deforming polynomial
$\xi_{\ell}(\eta;\bm{\lambda})$ will be given in \eqref{QMxil}.
For even $\ell$, the polynomial $\xi_{\ell}(\eta(x);\bm{\lambda})$ has
no zero in the range of $x$ and the modified Hamiltonian system is
hermitian, which reads:
\begin{align}
  &\mathcal{A}_{\ell}(\bm{\lambda})\eqdef
  \frac{d}{dx}-\frac{dw_{\ell}(x;\bm{\lambda})}{dx},\quad
  \mathcal{A}_{\ell}(\bm{\lambda})^{\dagger}
  =-\frac{d}{dx}-\frac{dw_{\ell}(x;\bm{\lambda})}{dx},\\
  &\mathcal{H}_{\ell}(\bm{\lambda})\eqdef
  \mathcal{A}_{\ell}(\bm{\lambda})^{\dagger}
  \mathcal{A}_{\ell}(\bm{\lambda}),\\
  &\mathcal{H}_{\ell}(\bm{\lambda})\phi_{\ell,n}(x;\bm{\lambda})
  =\mathcal{E}_n(\bm{\lambda})\phi_{\ell,n}(x;\bm{\lambda})\quad
  (n=0,\ell+1,\ell+2,\ldots).
\end{align}
This system is not shape invariant.
As mentioned in section \ref{summary}, the above form of the deformed
prepotential \eqref{defellprep} is closely related to that of the
exceptional Laguerre and Jacobi polynomials.

A degree $\ell$ polynomial in $\eta$, $\xi_{\ell}(\eta;\bm{\lambda})$
is defined by
\begin{equation}
  \text{W}[P_1,\ldots,P_{\ell}](\eta;\bm{\lambda})
  \eqdef\prod_{k=1}^{\ell}k!\,c_k(\bm{\lambda})\cdot
  \xi_{\ell}(\eta;\bm{\lambda}),
\end{equation}
and the explicit forms are:
\begin{equation}
  \xi_{\ell}(\eta;\bm{\lambda})=\left\{
  \begin{array}{ll}
  \frac{1}{2^{\ell}\ell!\,i^{\ell}}\,H_{\ell}(i\eta)&:\text{H}\\
  L_{\ell}^{(-g-\ell-\frac12)}(-\eta)&:\text{L}\\
  \frac{(-2)^{\ell}}{(g+h+1)_{\ell}}
 P_{\ell}^{(-g-\ell-\frac12,-h-\ell-\frac12)}(\eta)&:\text{J}
  \end{array}\right.\!\!.
  \label{QMxil}
\end{equation}
The eigenfunctions are
\begin{align}
  &\phi_{\ell,0}(x;\bm{\lambda})\eqdef e^{w_{\ell}(x;\bm{\lambda})}
  =\frac{\phi_0(x;\bm{\lambda}+\ell\bm{\delta})}
  {\xi_{\ell}\bigl(\eta(x);\bm{\lambda}\bigr)},\quad
  \phi_{\ell,n}(x;\bm{\lambda})=\phi_{\ell,0}(x;\bm{\lambda})
  P_{\ell,n}\bigl(\eta(x);\bm{\lambda}\bigr),\\
  &\text{W}[P_1,\ldots,P_{\ell},P_n](\eta;\bm{\lambda})
  \eqdef\prod_{k=1}^{\ell}k!\,c_k(\bm{\lambda})\cdot(-1)^{\ell}
  P_{\ell,n}(\eta;\bm{\lambda})\quad
  \bigl(\Rightarrow P_{\ell,0}(\eta;\bm{\lambda})=1\bigr).
\end{align}
Note that $P_{\ell,n}(\eta;\bm{\lambda})$ is a polynomial of degree $n$
in $\eta$ and $P_{0,n}(\eta;\bm{\lambda})=P_n(\eta;\bm{\lambda})$ and
$P_{\ell,n}(\eta;\bm{\lambda})$$=0$ for $1\leq n\leq\ell$.
We set $P_{\ell,n}(\eta;\bm{\lambda})=0$ for $n<0$.
For even $\ell$, the eigenpolynomial $P_{\ell,n}(\eta(x);\bm{\lambda})$
($n\geq\ell+1$) has $n-\ell$ zeros in the range of $x$.
The operators $\mathcal{A}_{\ell}(\bm{\lambda})$ and
$\mathcal{A}_{\ell}(\bm{\lambda})^{\dagger}$ connect the modified system
$\mathcal{H}_\ell(\bm{\lambda})
=\mathcal{A}_{\ell}(\bm{\lambda})^{\dagger}\mathcal{A}_{\ell}(\bm{\lambda})$
to the shape invariant system
$\mathcal{H}_\ell'(\bm{\lambda})=
\mathcal{A}_{\ell}(\bm{\lambda})\mathcal{A}_{\ell}(\bm{\lambda})^{\dagger}$
with the parameters $\bm{\lambda}+(\ell+1)\bm{\delta}$,
which is denoted by the rightmost vertical line in Fig.\,2.
The $n$-th level ($n\geq\ell+1$) of the modified system $\mathcal{H}_\ell$
is {\em iso-spectral\/} with the $n-\ell-1$-th level of the new
shape invariant system $\mathcal{H}_\ell'$:
\begin{align}
  &\mathcal{A}_{\ell}(\bm{\lambda})\phi_{\ell,n}(x;\bm{\lambda})
  =f_{\ell,n}(\bm{\lambda})
  \phi_{n-\ell-1}\bigl(x;\bm{\lambda}+(\ell+1)\bm{\delta}\bigr),\\
  &\mathcal{A}_{\ell}(\bm{\lambda})^{\dagger}
  \phi_{n-\ell-1}\bigl(x;\bm{\lambda}+(\ell+1)\bm{\delta}\bigr)
  =b_{\ell,n-1}(\bm{\lambda})\phi_{\ell,n}(x;\bm{\lambda}).
\end{align}
Here $f_{\ell,n}(\bm{\lambda})$ and $b_{\ell,n-1}(\bm{\lambda})$ are
defined by
\begin{equation}
  \begin{array}{l}
  \ \ \,f_{\ell,n}(\bm{\lambda})\eqdef f_n(\bm{\lambda})\times A,\\[2pt]
  b_{\ell,n-1}(\bm{\lambda})\eqdef b_{n-1}(\bm{\lambda})\times A^{-1},
  \end{array}\quad
  A=\left\{
  \begin{array}{ll}
  (-2)^{\ell}(n-\ell)_{\ell}&:\text{H}\\
  1&:\text{L}\\
  (-2)^{-\ell}(n+g+h+1)_{\ell}&:\text{J}
  \end{array}\right.\!\!,
  \label{QMflnbln}
\end{equation}
and they factorise $\mathcal{E}_n(\bm{\lambda})$,
$\mathcal{E}_n(\bm{\lambda})
=f_{\ell,n}(\bm{\lambda})b_{\ell,n-1}(\bm{\lambda})$.
The forward and backward shift operators $\mathcal{F}_{\ell}(\bm{\lambda})$
and $\mathcal{B}_{\ell}(\bm{\lambda})$, which act on the polynomial
eigenfunctions, are defined by:
\begin{align}
  &\mathcal{F}_{\ell}(\bm{\lambda})\eqdef
  \phi_0\bigl(x;\bm{\lambda}+(\ell+1)\bm{\delta}\bigr)^{-1}\circ
  \mathcal{A}_{\ell}(\bm{\lambda})\circ\phi_{\ell,0}(x;\bm{\lambda})\n
  &\phantom{\mathcal{F}_{\ell}(\bm{\lambda})}
  =\frac{\phi_0(x;\bm{\lambda}+\ell\bm{\delta})}
  {\phi_0(x;\bm{\lambda}+(\ell+1)\bm{\delta})}
  \frac{1}{\xi_{\ell}(\eta(x);\bm{\lambda})}\,\frac{d}{dx}\,,\\
  &\mathcal{B}_{\ell}(\bm{\lambda})\eqdef
  \phi_{\ell,0}(x;\bm{\lambda})^{-1}\circ
  \mathcal{A}_{\ell}(\bm{\lambda})^{\dagger}
  \circ\phi_0\bigl(x;\bm{\lambda}+(\ell+1)\bm{\delta}\bigr)\n
  &\phantom{\mathcal{B}_{\ell}(\bm{\lambda})}
  =-\frac{\phi_0(x;\bm{\lambda}+(\ell+1)\bm{\delta})}
  {\phi_0(x;\bm{\lambda}+\ell\bm{\delta})}\,
  \xi_{\ell}(\eta(x);\bm{\lambda})\n
  &\qquad\qquad\times
  \Bigl(\frac{d}{dx}+\partial_x\bigl(
  \mathcal{W}(x;\bm{\lambda}+\ell\bm{\delta})
  +\mathcal{W}(x;\bm{\lambda}+(\ell+1)\bm{\delta})\bigr)
  -\frac{\partial_x\xi_{\ell}(\eta(x);\bm{\lambda})}
  {\xi_{\ell}(\eta(x);\bm{\lambda})}\Bigr).
\end{align}
Their action on the polynomials is ($n\geq\ell+1$):
\begin{align}
  &\mathcal{F}_{\ell}(\bm{\lambda})P_{\ell,n}(\eta(x);\bm{\lambda})
  =f_{\ell,n}(\bm{\lambda})
  P_{n-\ell-1}\bigl(\eta(x);\bm{\lambda}+(\ell+1)\bm{\delta}\bigr),\\
  &\mathcal{B}_{\ell}(\bm{\lambda})
  P_{n-\ell-1}\bigl(\eta(x);\bm{\lambda}+(\ell+1)\bm{\delta}\bigr)
  =b_{\ell,n-1}(\bm{\lambda})P_{\ell,n}(\eta(x);\bm{\lambda}).
  \label{backpln}
\end{align}
Note that $\mathcal{F}_{\ell}(\bm{\lambda})$ and
$\mathcal{B}_{\ell}(\bm{\lambda})$ can be expressed in terms of $\eta$
\cite{hos}.
For $n\geq\ell+1$, the above relation \eqref{backpln} provides a simple
formula of the modified eigenpolynomial $P_{\ell,n}(\eta;\bm{\lambda})$
in terms of $\xi_{\ell}(\eta;\bm{\lambda})$ and the original eigenpolynomial
$P_{n}(\eta;\bm{\lambda})$:
\begin{align}
  &\quad b_{\ell,n-1}(\bm{\lambda})f_{n-\ell}(\bm{\lambda}+\ell\bm{\delta})
  P_{\ell,n}(\eta;\bm{\lambda})\n
  &=\mathcal{E}_{n-\ell}(\bm{\lambda}+\ell\bm{\delta})
  \xi_{\ell}(\eta;\bm{\lambda})P_{n-\ell}(\eta;\bm{\lambda}+\ell\bm{\delta})
  +4c_2(\eta)\partial_{\eta}\xi_{\ell}(\eta;\bm{\lambda})
  \,\partial_{\eta}P_{n-\ell}(\eta;\bm{\lambda}+\ell\bm{\delta}),
\end{align}
in which the coefficient $c_2(\eta)$ is given by
\begin{equation}
  c_2(\eta)\eqdef\left\{
  \begin{array}{ll}
  \frac14&:\text{H}\\
  \eta&:\text{L}\\
  1-\eta^2&:\text{J}
  \end{array}\right.\!\!.
\end{equation}
The orthogonality relation for even $\ell$ is:
\begin{align}
  &\int_{x_1}^{x_2}\!\!\phi_{\ell,0}(x;\bm{\lambda})^2\,
  P_{\ell,n}(\eta(x);\bm{\lambda})P_{\ell,m}(\eta(x);\bm{\lambda})dx
  =h_{\ell,n}(\bm{\lambda})\delta_{nm},\\
  &h_{\ell,n}(\bm{\lambda})\eqdef
  (n-\ell)_{\ell}\,h_n(\bm{\lambda})\times\left\{
  \begin{array}{ll}
  2^{\ell}&:\text{H}\\
  1&:\text{L}\\
  4^{-\ell}(n+g+h+1)_{\ell}&:\text{J}
  \end{array}\right.\!\!,
  \ \ (n=0,\ n\geq\ell+1).
\end{align}

\bigskip
A few historical remarks are in order.
Dubov et al \cite{dubov} derived in 1992 an exactly solvable Hamiltonian
system of a deformed harmonic oscillator, which corresponds to the
$\ell=2$ case of this Appendix. Their paper, written about two years
before Adler's, relied on rather heuristic arguments.
Recently Quesne \cite{quesne2} derived exactly solvable and non-shape
invariant systems of deformed radial oscillator and deformed DPT
potential, both are called type III. These results again correspond to
the $\ell=2$ cases of the radial oscillator and the DPT potential of
this Appendix.

\section{The discrete QM}
\setcounter{equation}{0}

Here we apply Adler's theorem to the shape invariant, therefore solvable,
systems whose eigenfunctions are described by the orthogonal polynomials;
the Meixner-Pollaczek (we set the parameter $\phi=\frac{\pi}{2}$),
continuous Hahn, Wilson and Askey-Wilson polynomials \cite{koeswart},
to be abbreviated as MP, cH, W and AW, respectively. See \cite{os4} and
\cite{os13} for the discrete QM treatment of these polynomials.

\subsection{The original systems}

Here we summarise various properties of the original Hamiltonian systems
to be compared with the specially modified systems to be presented in
\ref{sec:B2}.
Let us start with the Hamiltonians, Schr\"{o}dinger equations and
eigenfunctions ($x_1<x<x_2$):
\begin{align}
  &\mathcal{A}(\bm{\lambda})\eqdef
  i\bigl(e^{\frac{\gamma}{2}p}\sqrt{V^*(x;\bm{\lambda})}
  -e^{-\frac{\gamma}{2}p}\sqrt{V(x;\bm{\lambda})}\,\bigr),\n
  &\mathcal{A}(\bm{\lambda})^{\dagger}\eqdef
  -i\bigl(\sqrt{V(x;\bm{\lambda})}\,e^{\frac{\gamma}{2}p}
  -\sqrt{V^*(x;\bm{\lambda})}\,e^{-\frac{\gamma}{2}p}\bigr),\\
  &\mathcal{H}(\bm{\lambda})\eqdef
  \mathcal{A}(\bm{\lambda})^{\dagger}\mathcal{A}(\bm{\lambda}),\\
  &\mathcal{H}(\bm{\lambda})\phi_n(x;\bm{\lambda})
  =\mathcal{E}_n(\bm{\lambda})\phi_n(x;\bm{\lambda})\quad
  (n=0,1,2,\ldots),\\
  &\phi_n(x;\bm{\lambda})=\phi_0(x;\bm{\lambda})P_n(\eta(x);\bm{\lambda}).
\end{align}
The set of parameters $\bm{\lambda}$ are
\begin{align}
  \text{MP}:\quad&\bm{\lambda}\eqdef a,\quad a>0,\\
  \text{cH}:\quad&\bm{\lambda}\eqdef(a_1,a_2),\quad
  \text{Re}\,{a_i}>0 \ \ (i=1,2),\\
  \text{W}:\quad&\bm{\lambda}\eqdef(a_1,a_2,a_3,a_4),\quad
  \text{Re}\,a_i>0\ \ (i=1,\ldots,4),\n
  &\qquad
  \{a_1^*,a_2^*,a_3^*,a_4^*\}=\{a_1,a_2,a_3,a_4\} \quad (\text{as a set}),\\
  \text{AW}:\quad&q^{\bm{\lambda}}\eqdef(a_1,a_2,a_3,a_4),\quad
  |a_i|<1,\ \ (i=1,\ldots,4),\quad 0<q<1,\n
  &\qquad
  \{a_1^*,a_2^*,a_3^*,a_4^*\}=\{a_1,a_2,a_3,a_4\}\ \ (\text{as a set}),
\end{align}
where
$q^{(\lambda_1,\lambda_2,\ldots)}\eqdef(q^{\lambda_1},q^{\lambda_2},\ldots)$.
The the sinusoidal coordinate  $\eta(x)$ is,
\begin{equation}
  \eta(x)\eqdef\left\{
  \begin{array}{lllll}
  x,&x_1=-\infty,& x_2=\infty,&\gamma=1&:\text{MP,\,cH}\\
  x^2,&x_1=0,&x_2=\infty,&\gamma=1&:\text{W}\\
  \cos x,&x_1=0,& x_2=\pi,&\gamma=\log q&:\text{AW}
  \end{array}\right.\!\!.
  \label{dsinu}
\end{equation}
The potential function $V(x;\bm{\lambda})$ and energy eigenvalue
$\mathcal{E}_n(\bm{\lambda})$ are
\begin{align}
  &V(x;\bm{\lambda})\eqdef\left\{
  \begin{array}{ll}
  a+ix&:\text{MP}\\
  (a_1+ix)(a_2+ix)&:\text{cH}\\[2pt]
  \bigl(2ix(2ix+1)\bigr)^{-1}\prod_{j=1}^4(a_j+ix)&:\text{W}\\[2pt]
  \bigl((1-e^{2ix})(1-qe^{2ix})\bigr)^{-1}\prod_{j=1}^4(1-a_je^{ix})
  &:\text{AW}
  \end{array}\right.\!\!,
  \label{Vform}\\
  &\mathcal{E}_n(\bm{\lambda})\eqdef\left\{
  \begin{array}{lll}
  2n&&:\text{MP}\\
  n(n+b_1-1),&b_1\eqdef a_1+a_2+a_1^*+a_2^*&:\text{cH}\\[1pt]
  n(n+b_1-1),&b_1\eqdef a_1+a_2+a_3+a_4&:\text{W}\\[1pt]
  (q^{-n}-1)(1-b_4q^{n-1}),&b_4\eqdef a_1a_2a_3a_4&:\text{AW}
  \end{array}\right.\!\!.
\end{align}
The eigenfunction is described by the orthogonal polynomial
$P_n(\eta;\bm{\lambda})$, a polynomial of degree $n$ in $\eta$:
\begin{align}
  &P_n(\eta;\bm{\lambda})\eqdef c_n(\bm{\lambda})
  P_n^{\text{monic}}(\eta;\bm{\lambda}),\\
  &P_n(\eta;\bm{\lambda})\eqdef\left\{
  \begin{array}{ll}
  P_n^{(a)}(\eta;\tfrac{\pi}{2})\!\!&:\text{MP}\\
  p_n(\eta;a_1,a_2,a_1^*,a_2^*)\!\!&:\text{cH}\\[1pt]
  W_n(\eta;a_1,a_2,a_3,a_4)\!\!&:\text{W}\\[1pt]
  p_n(\eta;a_1,a_2,a_3,a_4|q)\!\!&:\text{AW}
  \end{array}\right.\!\!,
  \ \ c_n(\bm{\lambda})\eqdef\left\{
  \begin{array}{ll}
  \frac{1}{n!}\,2^n\!\!&:\text{MP}\\[1pt]
  \frac{1}{n!}\,(n+b_1-1)_n\!\!&:\text{cH}\\[1pt]
  (-1)^n(n+b_1-1)_n\!\!&:\text{W}\\[1pt]
  2^n(b_4q^{n-1}\,;q)_n\!\!&:\text{AW}
  \end{array}\right.\!\!,
\end{align}
in which $(a;q)_n$ is the $q$-Pochhammer symbol.
We set $P_n(\eta;\bm{\lambda})=0$ for $n<0$.
The shape invariance relations involve one more parameter $\kappa$:
\begin{align}
  &\mathcal{A}(\bm{\lambda})\mathcal{A}(\bm{\lambda})^{\dagger}
  =\kappa\mathcal{A}(\bm{\lambda+\bm{\delta}})^{\dagger}
  \mathcal{A}(\bm{\lambda}+\bm{\delta})
  +\mathcal{E}_1(\bm{\lambda}),\\
  &\quad\bm{\delta}\eqdef\left\{
  \begin{array}{ll}
  \frac12&:\text{MP}\\[1pt]
  (\frac12,\frac12)&:\text{cH}\\[1pt]
  (\frac12,\frac12,\frac12,\frac12)&:\text{W,\,AW}
  \end{array}\right.\!\!,
  \quad
  \kappa\eqdef\left\{
  \begin{array}{ll}
  1&:\text{MP,\,cH,\,W}\\
  q^{-1}&:\text{AW}
  \end{array}\right.\!\!,
\end{align}
or equivalently,
\begin{align}
  V(x-i\tfrac{\gamma}{2};\bm{\lambda})
  V^*(x-i\tfrac{\gamma}{2};\bm{\lambda})
  &=\kappa^2\,V(x;\bm{\lambda}+\bm{\delta})
  V^*(x-i\gamma;\bm{\lambda}+\bm{\delta}),\\
  V(x+i\tfrac{\gamma}{2};\bm{\lambda})
  +V^*(x-i\tfrac{\gamma}{2};\bm{\lambda})
  &=\kappa\bigl(V(x;\bm{\lambda}+\bm{\delta})
  +V^*(x;\bm{\lambda}+\bm{\delta})\bigr)
  -\mathcal{E}_1(\bm{\lambda}).
\end{align}
The groundstate wavefunction $\phi_0(x;\bm{\lambda})$ is determined by
\begin{equation}
  \sqrt{V^*(x-i\tfrac{\gamma}{2};\bm{\lambda})}
  \,\phi_0(x-i\tfrac{\gamma}{2};\bm{\lambda})
  =\sqrt{V(x+i\tfrac{\gamma}{2};\bm{\lambda})}
  \,\phi_0(x+i\tfrac{\gamma}{2};\bm{\lambda}),
\end{equation}
and its explicit forms are:
\begin{equation}
  \phi_0(x;\bm{\lambda})\eqdef\left\{
  \begin{array}{ll}
  \sqrt{\Gamma(a+ix)\Gamma(a-ix)}&:\text{MP}\\[2pt]
  \sqrt{\Gamma(a_1+ix)\Gamma(a_2+ix)\Gamma(a_1^*-ix)\Gamma(a_2^*-ix)}
  &:\text{cH}\\[2pt]
  \sqrt{(\Gamma(2ix)\Gamma(-2ix))^{-1}
  \prod_{j=1}^4\Gamma(a_j+ix)\Gamma(a_j-ix)}&:\text{W}\\[2pt]
  \sqrt{(e^{2ix}\,;q)_{\infty}(e^{-2ix}\,;q)_{\infty}
  \prod_{j=1}^4(a_je^{ix}\,;q)_{\infty}^{-1}(a_je^{-ix}\,;q)_{\infty}^{-1}}
  &:\text{AW}
  \end{array}\right.\!\!.
\end{equation}
We introduce an auxiliary function $\varphi(x)$ with the properties:
\begin{align}
  &\varphi(x)\eqdef\left\{
  \begin{array}{ll}
  1&:\text{MP,\,cH}\\
  2x&:\text{W}\\
  2\sin x&:\text{AW}
  \end{array}\right.\!\!,
  \label{varphidef}\\
  &\phi_0(x;\bm{\lambda}+\bm{\delta})
  =\varphi(x)\sqrt{V(x+i\tfrac{\gamma}{2};\bm{\lambda})}\,
  \phi_0(x+i\tfrac{\gamma}{2};\bm{\lambda}),\\
  &V(x;\bm{\lambda}+\bm{\delta})
  =\kappa^{-1}\frac{\varphi(x-i\gamma)}{\varphi(x)}
  V(x-i\tfrac{\gamma}{2};\bm{\lambda}).
  \label{varphiprop3}
\end{align}
The sinusoidal coordinate $\eta(x)$ has the following properties:
\begin{align}
  &\eta(x-i\tfrac{k\gamma}{2})-\eta(x+i\tfrac{k\gamma}{2})
  =-i\varphi(x)\times\left\{
  \begin{array}{ll}
  k&:\text{MP,\,cH,\,W}\\
  \sinh\frac{-k\gamma}{2}&:\text{AW}
  \end{array}\right.\!\!,\\
  &\eta(x-i\tfrac{k\gamma}{2})+\eta(x+i\tfrac{k\gamma}{2})
  =\left\{
  \begin{array}{ll}
  2\eta(x)&:\text{MP,\,cH}\\
  2\eta(x)-\tfrac12k^2&:\text{W}\\[1pt]
  2\eta(x)\cosh\tfrac{k\gamma}{2}&:\text{AW}
  \end{array}\right.\!\!,\\
  &\eta(x-i\tfrac{k\gamma}{2})\eta(x+i\tfrac{k\gamma}{2})
  =\left\{
  \begin{array}{ll}
  \eta(x)^2+\tfrac14k^2&:\text{MP,\,cH}\\
  \bigl(\eta(x)+\tfrac14k^2\bigr)^2&:\text{W}\\[1pt]
  \eta(x)^2+\sinh^2\tfrac{k\gamma}{2}&:\text{AW}
  \end{array}\right.\!\!.
  \label{etaprod}
\end{align}
These mean that for a polynomial $P(\eta)$ in $\eta$,
$i\varphi(x)^{-1}\bigl(P(\eta(x-i\frac{k\gamma}{2}))
-P(\eta(x+i\frac{k\gamma}{2}))\bigr)$ is another polynomial in $\eta(x)$.
The action of $\mathcal{A}(\bm{\lambda})$ and
$\mathcal{A}(\bm{\lambda})^{\dagger}$ on the eigenfunctions is
\begin{equation}
  \mathcal{A}(\bm{\lambda})\phi_n(x;\bm{\lambda})
  =f_n(\bm{\lambda})
  \phi_{n-1}\bigl(x;\bm{\lambda}+\bm{\delta}\bigr),\quad
  \mathcal{A}(\bm{\lambda})^{\dagger}
  \phi_{n-1}\bigl(x;\bm{\lambda}+\bm{\delta}\bigr)
  =b_{n-1}(\bm{\lambda})\phi_n(x;\bm{\lambda}).
\end{equation}
The factors of the energy eigenvalue, $f_n(\bm{\lambda})$ and
$b_{n-1}(\bm{\lambda})$,
$\mathcal{E}_n(\bm{\lambda})=f_n(\bm{\lambda})b_{n-1}(\bm{\lambda})$,
are given by
\begin{equation}
  f_n(\bm{\lambda})\eqdef\left\{
  \begin{array}{ll}
  2&:\text{MP}\\
  n+b_1-1&:\text{cH}\\
  -n(n+b_1-1)&:\text{W}\\
  q^{\frac{n}{2}}(q^{-n}-1)(1-b_4q^{n-1})&:\text{AW}
  \end{array}\right.\!\!,
  \quad
  b_{n-1}(\bm{\lambda})\eqdef\left\{
  \begin{array}{ll}
  n&:\text{MP,\,cH}\\
  -1&:\text{W}\\
  q^{-\frac{n}{2}}&:\text{AW}
  \end{array}\right.\!\!.
\end{equation}
The forward and backward shift operators $\mathcal{F}(\bm{\lambda})$ and
$\mathcal{B}(\bm{\lambda})$ are defined by
\begin{align}
  &\mathcal{F}(\bm{\lambda})\eqdef
  \phi_0(x;\bm{\lambda}+\bm{\delta})^{-1}\circ
  \mathcal{A}(\bm{\lambda})\circ\phi_0(x;\bm{\lambda})
  =i\varphi(x)^{-1}(e^{\frac{\gamma}{2}p}-e^{-\frac{\gamma}{2}p}),\\
  &\mathcal{B}(\bm{\lambda})\eqdef
  \phi_0(x;\bm{\lambda})^{-1}\circ
  \mathcal{A}(\bm{\lambda})^{\dagger}
  \circ\phi_0(x;\bm{\lambda}+\bm{\delta})
  =-i\bigl(V(x;\bm{\lambda})e^{\frac{\gamma}{2}p}
  -V^*(x;\bm{\lambda})e^{-\frac{\gamma}{2}p}\bigr)\varphi(x),
\end{align}
and their action on the polynomials is
\begin{align}
  &\mathcal{F}(\bm{\lambda})P_n(\eta(x);\bm{\lambda})
  =f_n(\bm{\lambda})P_{n-1}(\eta(x);\bm{\lambda}+\bm{\delta}),\\
  &\mathcal{B}(\bm{\lambda})P_{n-1}(\eta(x);\bm{\lambda}+\bm{\delta})
  =b_{n-1}(\bm{\lambda})P_n(\eta(x);\bm{\lambda}).
\end{align}
The orthogonality relation is
\begin{align}
  &\int_{x_1}^{x_2}\!\!\phi_0(x;\bm{\lambda})^2\,
  P_n(\eta(x);\bm{\lambda})P_m(\eta(x);\bm{\lambda})dx
  =h_n(\bm{\lambda})\delta_{nm},\\
  &h_n(\bm{\lambda})\eqdef\left\{
  \begin{array}{ll}
  2\pi\,\Gamma(n+2a)\bigl(n!\,2^{2a}\bigr)^{-1}&:\text{MP}\\
  2\pi\prod_{i,j=1}^2\Gamma(n+a_i+a_j^*)\cdot
  \bigl(n!(2n+b_1-1)\Gamma(n+b_1-1)\bigr)^{-1}&:\text{cH}\\[2pt]
  2\pi n!\,(n+b_1-1)_n\prod_{1\leq i<j\leq 4}\Gamma(n+a_i+a_j)\cdot
  \Gamma(2n+b_1)^{-1}&:\text{W}\\[2pt]
  2\pi(b_4q^{n-1};q)_n(b_4q^{2n};q)_{\infty}(q^{n+1};q)_{\infty}^{-1}
  \prod_{1\leq i<j\leq 4}(a_ia_jq^n;q)_{\infty}^{-1}&:\text{AW}
  \end{array}\right.\!\!.
\end{align}

\subsection{The $(\phi_1,\ldots,\phi_{\ell})$-deleted systems}
\label{sec:B2}

The potential function, the Hamiltonian and the Schr\"{o}dinger equation
of the modified system are:
\begin{align}
  &V_{\ell}(x;\bm{\lambda})\eqdef
  \frac{\varphi(x-i\frac{\ell+1}{2}\gamma)\varphi(x-i\frac{\ell}{2}\gamma)}
  {\varphi(x)\varphi(x-i\frac{\gamma}{2})}
  \frac{\xi_{\ell}(\eta(x+i\frac{\gamma}{2});\bm{\lambda})}
  {\xi_{\ell}(\eta(x-i\frac{\gamma}{2});\bm{\lambda})}
  V(x-i\tfrac{\ell}{2}\gamma;\bm{\lambda})
  \label{vellexp1}\\
  &\phantom{V_{\ell}(x;\bm{\lambda})}=\kappa^\ell
  \,\frac{\xi_{\ell}(\eta(x+i\frac{\gamma}{2});\bm{\lambda})}
  {\xi_{\ell}(\eta(x-i\frac{\gamma}{2});\bm{\lambda})}
  V(x;\bm{\lambda}+\ell\bm{\delta}),
  \label{vellexp2}\\
  &\mathcal{A}_{\ell}(\bm{\lambda})\eqdef
   i\bigl(e^{\frac{\gamma}{2}p}\sqrt{V_{\ell}^*(x;\bm{\lambda})}
  -e^{-\frac{\gamma}{2}p}\sqrt{V_{\ell}(x;\bm{\lambda})}\,\bigr),\n
  &\mathcal{A}_{\ell}(\bm{\lambda})^{\dagger}\eqdef
   -i\bigl(\sqrt{V_{\ell}(x;\bm{\lambda})}\,e^{\frac{\gamma}{2}p}
  -\sqrt{V_{\ell}^*(x;\bm{\lambda})}\,e^{-\frac{\gamma}{2}p}\bigr),\\
  &\mathcal{H}_{\ell}(\bm{\lambda})\eqdef
  \mathcal{A}_{\ell}(\bm{\lambda})^{\dagger}
  \mathcal{A}_{\ell}(\bm{\lambda}),\\
  &\mathcal{H}_{\ell}(\bm{\lambda})\phi_{\ell,n}(x;\bm{\lambda})
  =\mathcal{E}_n(\bm{\lambda})\phi_{\ell,n}(x;\bm{\lambda})\quad
  (n=0,\ell+1,\ell+2,\ldots).
\end{align}
The explicit forms of the deforming polynomial
$\xi_{\ell}(\eta;\bm{\lambda})$ will be given in \eqref{dQMxil}.
For even $\ell$, $\xi_{\ell}(\eta(x);\bm{\lambda})$ has no zero in the
rectangular domain in the complex $x$ plane,
$x_1\leq\text{Re}\,x\leq x_2$, $|\text{Im}\,x|\leq\frac12|\gamma|$.
Note that the normalization of $\xi_{\ell}$ does not affect
$\mathcal{H}_{\ell}$. This system is not shape invariant.
The second line of the expression for $V_{\ell}(x;\bm{\lambda})$,
\eqref{vellexp2}, is obtained from the first line \eqref{vellexp1} by
repeated use of the formula \eqref{varphiprop3} of the auxiliary
function $\varphi$. As mentioned in section \ref{summary}, this form
of the deformed potential function \eqref{vellexp2} is closely related
to that of the exceptional Wilson and Askey-Wilson polynomials.

It is convenient to introduce an auxiliary function $\varphi_{\ell}(x)$:
\begin{equation}
  \varphi_{\ell}(x)\eqdef
  \varphi(x)^{[\frac{\ell}{2}]}\prod_{k=1}^{\ell-2}
  \bigl(\varphi(x-i\tfrac{k}{2}\gamma)\varphi(x+i\tfrac{k}{2}\gamma)
  \bigr)^{[\frac{\ell-k}{2}]},
  \label{varphildef}
\end{equation}
where $[x]$ denotes the greatest integer not exceeding $x$.
Note that $[\frac{\ell}{2}]+2\sum_{k=1}^{\ell-2}[\frac{\ell-k}{2}]
=\frac12\ell(\ell-1)$.
The deforming polynomial $\xi_{\ell}(\eta;\bm{\lambda})$ is defined by
\begin{align}
  &\quad
  \text{W}_{\gamma}[\check{P}_1,\ldots,\check{P}_{\ell}](x;\bm{\lambda})
  \qquad\quad\bigl(
  \check{P}_n(x;\bm{\lambda})\eqdef P_n(\eta(x);\bm{\lambda})\bigr)\n
  &\eqdef\prod_{k=1}^{\ell}c_k(\bm{\lambda})\cdot
  \varphi_{\ell}(x)
  \xi_{\ell}(\eta(x);\bm{\lambda})
  \times\left\{
  \begin{array}{ll}
  \prod_{k=1}^{\ell}k!\!\!&:\text{MP,\,cH,\,W}\\
  \prod_{k=1}^{\ell}\prod_{j=1}^k\sinh\frac{-j\gamma}{2}\!\!&:\text{AW}
  \end{array}\right.\!\!.
\end{align}
As in the ordinary QM cases \eqref{QMxil}, it is expressed in terms of
the polynomial $P_\ell$ of the original system with shifted parameters:
\begin{equation}
  \xi_{\ell}(\eta;\bm{\lambda})=
  \frac{P_{\ell}(\eta;-\bm{\lambda}^*-(\ell-1)\bm{\delta})}
  {c_{\ell}(-\bm{\lambda}^*-(\ell-1)\bm{\delta})}
  \times\left\{
  \begin{array}{ll}
  (\ell!)^{-1}&:\text{MP,\,cH,\,W}\\
  (\prod_{j=1}^{\ell}\sinh\frac{-j\gamma}{2})^{-1}&:\text{AW}
  \end{array}\right.\!\!.
  \label{dQMxil}
\end{equation}
Note that $P_n(\eta;\bm{\lambda}^*)=P_n(\eta;\bm{\lambda})$
and $c_n(\bm{\lambda}^*)=c_n(\bm{\lambda})$ for the MP, W and AW cases.
The eigenfunctions are
\begin{align}
  &\phi_{\ell,0}(x;\bm{\lambda})
  \eqdef\frac{(-1)^{\ell}\kappa^{\frac14\ell(\ell-1)}
  \phi_0(x;\bm{\lambda}+\ell\bm{\delta})}
  {\sqrt{\xi_{\ell}(\eta(x-i\frac{\gamma}{2});\bm{\lambda})
  \xi_{\ell}(\eta(x+i\frac{\gamma}{2});\bm{\lambda})}},
  \ \,\phi_{\ell,n}(x;\bm{\lambda})=\phi_{\ell,0}(x;\bm{\lambda})
  P_{\ell,n}\bigl(\eta(x);\bm{\lambda}\bigr),\!\!\\
  &\text{W}_{\gamma}
  [\check{P}_1,\ldots,\check{P}_{\ell},\check{P}_n](x;\bm{\lambda})
  \eqdef\prod_{k=1}^{\ell}c_k(\bm{\lambda})\cdot
  \varphi_{\ell+1}(x)(-1)^{\ell}
  P_{\ell,n}(\eta(x);\bm{\lambda})\\
  &\qquad\qquad\qquad\qquad\qquad\qquad
  \times\left\{
  \begin{array}{ll}
  \prod_{k=1}^{\ell}k!&\!\!:\text{MP,\,cH,\,W}\\
  \prod_{k=1}^{\ell}\prod_{j=1}^k\sinh\frac{-j\gamma}{2}&\!\!:\text{AW}
  \end{array}\right.
  \bigl(\Rightarrow P_{\ell,0}(\eta;\bm{\lambda})=1\bigr).
  \nonumber
\end{align}
For even $\ell$, $P_{\ell,n}(\eta(x);\bm{\lambda})$ ($n\geq\ell+1$)
has $n-\ell$ zeros in the range of $x$.
Note that $P_{\ell,n}(\eta;\bm{\lambda})$ is a polynomial of degree $n$
in $\eta$ and $P_{0,n}(\eta;\bm{\lambda})=P_n(\eta;\bm{\lambda})$ and
$P_{\ell,n}(\eta;\bm{\lambda})=0$ for $1\leq n\leq\ell$.
We set $P_{\ell,n}(\eta;\bm{\lambda})=0$ for $n<0$.
The operators $\mathcal{A}_{\ell}(\bm{\lambda})$ and
$\mathcal{A}_{\ell}(\bm{\lambda})^{\dagger}$ connect the modified system
$\mathcal{H}_\ell(\bm{\lambda})=\mathcal{A}_{\ell}(\bm{\lambda})^{\dagger}
\mathcal{A}_{\ell}(\bm{\lambda})$
to the shape invariant system
$\mathcal{H}_\ell'(\bm{\lambda})=\mathcal{A}_{\ell}(\bm{\lambda})
\mathcal{A}_{\ell}(\bm{\lambda})^{\dagger}$ with the parameters
$\bm{\lambda}+(\ell+1)\bm{\delta}$,
which is denoted by the rightmost vertical line in Fig.\,2.
The $n$-th level ($n\geq\ell+1$) of the modified system $\mathcal{H}_\ell$
is {\em iso-spectral\/} with the $n-\ell-1$-th level of the new
shape invariant system $\mathcal{H}_\ell'$:
\begin{align}
  &\mathcal{A}_{\ell}(\bm{\lambda})\phi_{\ell,n}(x;\bm{\lambda})
  =f_{\ell,n}(\bm{\lambda})
  \phi_{n-\ell-1}\bigl(x;\bm{\lambda}+(\ell+1)\bm{\delta}\bigr),\\
  &\mathcal{A}_{\ell}(\bm{\lambda})^{\dagger}
  \phi_{n-\ell-1}\bigl(x;\bm{\lambda}+(\ell+1)\bm{\delta}\bigr)
  =b_{\ell,n-1}(\bm{\lambda})\phi_{\ell,n}(x;\bm{\lambda}).
\end{align}
Here, $f_{\ell,n}(\bm{\lambda})$ and $b_{\ell,n-1}(\bm{\lambda})$ are the
factors of the energy eigenvalue, $\mathcal{E}_n(\bm{\lambda})
=f_{\ell,n}(\bm{\lambda})b_{\ell,n-1}(\bm{\lambda})$, and are defined by
\begin{equation}
  \begin{array}{l}
  \ \ \,f_{\ell,n}(\bm{\lambda})\eqdef f_n(\bm{\lambda})\times A,\\[2pt]
  b_{\ell,n-1}(\bm{\lambda})\eqdef b_{n-1}(\bm{\lambda})\times A^{-1},
  \end{array}\quad
  A=\left\{
  \begin{array}{ll}
  2^{\ell}&:\text{MP}\\
  (b_1+n)_{\ell}&:\text{cH}\\
  (-1)^{\ell}(n-\ell)_{\ell}(b_1+n)_{\ell}&:\text{W}\\
  q^{-\frac12\ell n}(q^{n-\ell};q)_{\ell}(b_4q^n;q)_{\ell}&:\text{AW}
  \end{array}\right.\!\!.
  \label{dQMflnbln}
\end{equation}
The forward and backward shift operators $\mathcal{F}_{\ell}(\bm{\lambda})$
and $\mathcal{B}_{\ell}(\bm{\lambda})$ which act on the polynomial
eigenfunctions, are defined by:
\begin{align}
  &\mathcal{F}_{\ell}(\bm{\lambda})\eqdef
  \phi_0\bigl(x;\bm{\lambda}+(\ell+1)\bm{\delta}\bigr)^{-1}\circ
  \mathcal{A}_{\ell}(\bm{\lambda})\circ\phi_{\ell,0}(x;\bm{\lambda})
  =\frac{(-1)^{\ell}\kappa^{\frac14\ell(\ell+1)}}
  {\varphi(x)\xi_{\ell}(\eta(x);\bm{\lambda})}\,
  i\bigl(e^{\frac{\gamma}{2}p}-e^{-\frac{\gamma}{2}p}\bigr),\!\!\!\\
  &\mathcal{B}_{\ell}(\bm{\lambda})\eqdef
  \phi_{\ell,0}(x;\bm{\lambda})^{-1}\circ
  \mathcal{A}_{\ell}(\bm{\lambda})^{\dagger}
  \circ\phi_0\bigl(x;\bm{\lambda}+(\ell+1)\bm{\delta}\bigr)\n
  &\phantom{\mathcal{B}_{\ell}(\bm{\lambda})}
  =(-1)^{\ell}\kappa^{-\frac14\ell(\ell-3)}(-i)\Bigl(
  V(x;\bm{\lambda}+\ell\bm{\delta})
  \xi_{\ell}(\eta(x+i\tfrac{\gamma}{2});\bm{\lambda})e^{\frac{\gamma}{2}p}\n
  &\phantom{\mathcal{B}_{\ell}(\bm{\lambda})
  =(-1)^{\ell}\kappa^{-\frac14\ell(\ell-3)}(-i)\Bigl(}
  -V^*(x;\bm{\lambda}+\ell\bm{\delta})
  \xi_{\ell}(\eta(x-i\tfrac{\gamma}{2});\bm{\lambda})e^{-\frac{\gamma}{2}p}
  \Bigr)\varphi(x),
\end{align}
and their action on the polynomials is
\begin{align}
  &\mathcal{F}_{\ell}(\bm{\lambda})P_{\ell,n}(\eta(x);\bm{\lambda})
  =f_{\ell,n}(\bm{\lambda})
  P_{n-\ell-1}\bigl(\eta(x);\bm{\lambda}+(\ell+1)\bm{\delta}\bigr),\\
  &\mathcal{B}_{\ell}(\bm{\lambda})
  P_{n-\ell-1}\bigl(\eta(x);\bm{\lambda}+(\ell+1)\bm{\delta}\bigr)
  =b_{\ell,n-1}(\bm{\lambda})P_{\ell,n}(\eta(x);\bm{\lambda}).
  \label{dQMback}
\end{align}
For $n\geq\ell+1$, the above formula \eqref{dQMback} provides a simple
formula of the modified eigenpolynomial $P_{\ell,n}(\eta;\bm{\lambda})$
in terms of $\xi_{\ell}(\eta;\bm{\lambda})$ and the original eigenpolynomial
$P_{n}(\eta;\bm{\lambda})$:
\begin{align}
  &\quad(-1)^{\ell}\kappa^{\frac14\ell(\ell-3)}
  b_{\ell,n-1}(\bm{\lambda})P_{\ell,n}(\eta;\bm{\lambda})\n
  &=-i\Bigl(
  V(x;\bm{\lambda}+\ell\bm{\delta})
  \xi_{\ell}(\eta(x+i\tfrac{\gamma}{2});\bm{\lambda})
  \varphi(x-i\tfrac{\gamma}{2})
  P_{n-\ell-1}(\eta(x-i\tfrac{\gamma}{2});\bm{\lambda}+(\ell+1)\bm{\delta})\n
  &\qquad
  -V^*(x;\bm{\lambda}+\ell\bm{\delta})
  \xi_{\ell}(\eta(x-i\tfrac{\gamma}{2});\bm{\lambda})
  \varphi(x+i\tfrac{\gamma}{2})
  P_{n-\ell-1}(\eta(x+i\tfrac{\gamma}{2});\bm{\lambda}+(\ell+1)\bm{\delta})
  \Bigr).
\end{align}
The orthogonality relation for even $\ell$ is:
\begin{align}
  &\int_{x_1}^{x_2}\!\!\phi_{\ell,0}(x;\bm{\lambda})^2\,
  P_{\ell,n}(\eta(x);\bm{\lambda})P_{\ell,m}(\eta(x);\bm{\lambda})dx
  =h_{\ell,n}(\bm{\lambda})\delta_{nm},\\
  &h_{\ell,n}(\bm{\lambda})\eqdef
  h_n(\bm{\lambda})\times\left\{
  \begin{array}{ll}
  (n-\ell)_{\ell}2^{\ell}&:\text{MP}\\
  (n-\ell)_{\ell}(b_1+n)_{\ell}&:\text{cH,W}\\
  q^{-\ell n}(q^{n-\ell};q)_{\ell}(b_4q^n;q)_{\ell}&:\text{AW}
  \end{array}\right.\!\!,
  \ \ (n=0,\ n\geq\ell+1).
\end{align}


\end{document}